\documentstyle[twocolumn,aps,epsf]{revtex}
\begin{document}

\def\r{{\bf{r}}}
\def\k{{\bf{k}}}
\def\K{{\bf{K}}}
\def\M{{\bf{M}}}
\def\q{{\bf{q}}}
\def\Q{{\bf{Q}}}
\def\T{{\bf{T}}}
\def\mT{{\mbox{T}}}
\def\G{{\bf{G}}}
\def\tk{\tilde{\bf{k}}}
\def\tK{\tilde{\bf{K}}}
\def\tq{\tilde{\bf{q}}}
\def\tQ{\tilde{\bf{Q}}}
\def\tchi{\tilde{\chi}}
\def\p{{\bf{p}}}
\def\P{{\bf{P}}}
\def\R{{\bf{R}}}
\def\J{{\bf{J}}}
\def\grad{{\nabla}}
\newcommand{\bgam}{{\bf{\Gamma}}}
\newcommand{\bchi}{{\bf{\chi}}}
\newcommand{\bSig}{{\bf{\Sigma}}}
\title{The Dynamical Cluster Approximation: Non--Local Dynamics of 
Correlated Electron Systems}
\author{M. H. Hettler$^{1,2}$,  M. Mukherjee$^1$, 
M. Jarrell$^1$, and H. R. Krishnamurthy $^{1,3}$}
\address{$^1$ Department of Physics, University of Cincinnati, Cincinnati,
OH 45221}
\address{$^2$ Materials Science Division, Argonne National Laboratory, Argonne, IL 60439}
\address{$^3$ Department of Physics, IISc, Bangalore 560012, India}
\date{\today}
\maketitle
\begin{abstract}
We recently introduced the dynamical cluster approximation(DCA), a new 
technique that includes short-ranged dynamical correlations in addition to 
the local dynamics of the dynamical mean field approximation while preserving 
causality.  The technique is based on an iterative self--consistency scheme 
on a finite size periodic cluster.  The dynamical 
mean field approximation (exact result) is obtained by taking the cluster 
to a single site (the thermodynamic limit).  Here, we  provide details of our 
method, explicitly show that it is causal, systematic, $\Phi$-derivable, and
that it becomes conserving as the cluster size increases. 
We demonstrate the DCA
by applying it to
a Quantum Monte Carlo and Exact Enumeration study of the 
two-dimensional Falicov-Kimball model.  The resulting spectral functions 
preserve causality, and the spectra and the CDW transition temperature 
converge quickly and systematically to the thermodynamic limit as the 
cluster size increases.
\end{abstract}

\narrowtext

\section {Introduction} 

Strongly correlated electron systems have been at the center of theoretical 
and experimental research interest for several decades.  This interest
was greatly intensified by the discovery of heavy fermion metals and 
superconductors, and recently of the high-$T_c$ superconductors. 
The observation of non-Fermi liquid behavior first in the Cuprates and
later even in some heavy fermion systems has given further impetus.
Away from a transition, these materials are characterized by short-ranged
dynamical correlations such as the local correlations responsible for
the Kondo effect.  In addition, the doped cuprates display short-ranged
antiferromagnetic dynamical correlations thought to be responsible for
pair formation.  Some of this physics is captured by the simplest models 
of strongly correlated electrons, such as
the Hubbard Model (HM) and the periodic Anderson model (PAM).  Despite the 
short range of the dynamical correlations and numerous sophisticated 
techniques introduced since the inception of the models, they remain unsolved.

However, recently Metzner and Vollhardt showed \cite{metzvoll} that these 
models undergo significant simplification in the limit of infinite 
dimensions, $D=\infty$. In this limit, provided the kinetic energy is
scaled as $1/\sqrt D$, the self energy and vertex functions may be taken to be 
purely local in space although they retain a nontrivial frequency dependence. 
Consequently, the HM and PAM can be mapped onto a self--consistently
embedded Anderson impurity problem; i.e., a single correlated site
subject to a self--consistently determined energy dependent hybridization 
with a conduction electron ``bath'' or ``host'' representing the
remaining sites of the lattice, or equivalently (on eliminating this 
bath), to a dynamical mean field \cite{jarrell,georges}.  The resulting 
dynamical mean-field theory (DMFT) is exact in infinite dimensions and has
been use to establish the thermodynamic properties and phase diagrams of 
these models using Quantum Monte Carlo (QMC) and other 
methods\cite{jarrell,jarrell2,lisa}.

A similar self-consistent  single site theory can be obtained by
{\em{assuming}} a purely local self energy (and vertex functions) even
in finite dimensions. This yields the natural mean field theory for correlated 
lattice systems and is called the dynamical mean field approximation (DMFA).
While it has been shown that this 
approximation captures many key features of strongly correlated systems
even in a finite-dimensional context, the DMFA has some obvious and significant
limitations.  For example, the only dynamical correlations present are those 
which may be properly treated on a single site.  Therefore, there are no 
non-local dynamical correlations.  These are necessary, for example, to
describe phases with explicitly non--local order parameters or those with
lower symmetry than the lattice, of which d--wave 
superconductivity is perhaps the most prominent example.
But even phases with local order parameters (e.g. commensurate magnetism) 
will certainly be affected by the non--local dynamical correlations (spin waves) 
neglected by the DMFA. In addition, as we show in this paper, the DMFA 
is not a conserving approximation, with violations of the Ward identity 
associated with current conservation (the equation of continuity)
for any $D$, including the limit $D\to\infty$.  

Consequently, there have been efforts to extend the DMFA by inclusion of 
non--local correlations, which would correspond to $1/D$--corrections 
to the self energy of the $D=\infty$ models\cite{peter1,avi}. These 
efforts have failed to construct a causal theory, one that 
preserves spectral weight and which retains positive semidefinite
spectral functions, out of non--local Green functions. 
Such violations of positivity have been seen explicitly and discussed 
in the work by van Dongen\cite{peter1}.  Even in the sophisticated 
$\Phi$-derivable technique developed by Schiller and Ingersent\cite{avi}, 
violations of the sum rules occurred for moderately large values of the 
interaction strength in the Falicov--Kimball model (FKM).

A different approach by Smith and Si\cite{smith-si} allows for the
incorporation of non--local interactions in the original Hamiltonian
(beyond the Hartree level) by rescaling them  with 
the same $1/\sqrt{D}$--factor in the limit $D=\infty$ as the
kinetic energy. The resulting self energy remains local, and the system
maps to a impurity model coupled to both a Fermionic bath
(the electrons on the host) as well as a Bosonic bath (the two--particle
interactions). While this approach is attractive we believe that
this  scaling is difficult to justify formally. 
In addition, since the resulting effective theory is still
a single site theory, it does not allow one to address
some of the problems discussed above.

In a recent paper\cite{fkm} we introduced the dynamical cluster 
approximation (DCA), an iterative self-consistency scheme on a finite size 
periodic cluster of size $N_c$.  It extends the 
DMFA through the inclusion of short-ranged dynamical correlations, remains 
fully causal, and restores the conservation laws of Ward \cite{ward} and Baym
\cite{baym} when the
cluster becomes large.  The essential approximation of the DCA is to take 
the irreducible self energy $\Sigma^c (\K,\omega)$ of the cluster as a good 
approximation to the self energy of the real system at 
the cluster momenta, $\K$.  When $N_c$, 
the number of cluster momenta in the first Brillouin zone 
is relatively small, this approximation can only be justified if the 
self energy of the real system is weakly momentum dependent. 
Such a weak momentum dependence is realized in high dimensions (there 
is {\it no} momentum dependence  in $D=\infty$).  Then, a coarse grid of 
$\K$ points is sufficient to capture all the short ranged (but non--local) 
dynamics. In low dimensions, the validity of the approximation is less 
clear.  However, in many correlated systems the momentum dependence of 
the self energy is less important than its frequency dependence.
In addition, because of the coupling of the cluster to a much 
larger host, the method allows for a systematic finite size study that
is likely to converge faster than standard methods like exact
diagonalization, lattice QMC and the fluctuation exchange approximation
(FLEX)\cite{bickers}

In this work we present the first detailed discussion of the DCA.  The paper 
is organized as follows:  First, we review the DMFA and discuss its 
limitations.  Then, we review the steps of the DCA and discuss the details 
of the formalism for the first time.  We then apply the DCA to the half-filled 
FKM using Quantum Monte Carlo and exact enumeration for the cluster problem 
to obtain self energies and Green functions. For simplicity, we consider only 
the single-band model with nearest-neighbor hopping on a periodic square 
lattice with $N$ sites.  We demonstrate that the DCA algorithm converges 
systematically with increasing cluster size and remains fully causal. We then 
discuss the results and their implications.  In the appendices, we provide 
the formalism needed to calculate the two-particle properties, generalize 
our formalism to models with extended range interactions, prove that it is 
causal, and discuss it conserving properties.

\section{Dynamical Mean Field Approximation}
\begin{figure}
\epsfxsize=3.6in
\epsffile{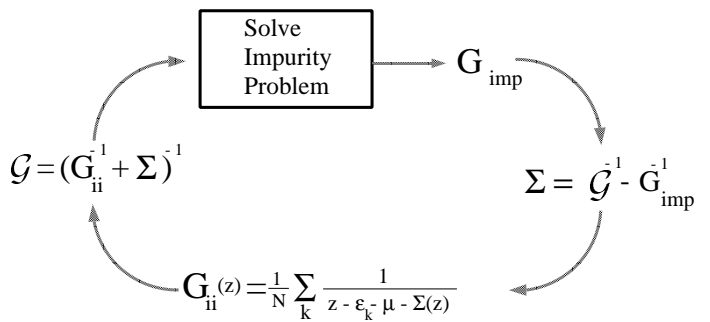}
\caption[a]{\em{Sketch of the DMFA algorithm.}}
\label{DMFA_alg}
\end{figure}

The DMFA\cite{metzvoll} may be derived in any dimension by disregarding 
momentum conservation at the internal vertices of the self 
energy\cite{muller-hartmann}.  This approximation
becomes exact in the limit of infinite dimensions $D\to\infty$,
provided that the near-neighbor
electronic hopping integral is rescaled so that $t\sim D^{-1/2}$.  Then,
the single-particle Green function $G(r)\sim t^{r}\sim D^{-r/2}$
and the self energy becomes a purely local functional of the local Green 
function only, $\Sigma_{i,j} = \Sigma_{i,i}(G_{i,i}) \delta_{i,j}$
which is momentum independent 
$\Sigma(\k,\omega)=\Sigma_{i,i}(\omega) +{\cal{O}}(1/\sqrt{D})$. The lattice 
problem may then be mapped onto a self-consistently embedded impurity problem.  
The resulting DMFA algorithm, illustrated in Fig.~\ref{DMFA_alg}, has 
the following steps:
(0) The procedure starts with a guess for $\Sigma_{ii}(\omega)$, usually
zero.
(1) Then, we calculate the local lattice Green function 
$G_{i,i}(\omega)={1\over N} \sum_{{\k}}(G_{o}^{-1}(\k,\omega) - 
\Sigma_{i,i}(\omega))^{-1}$, 
where $G_{o}(\k,\omega)$ is the bare lattice Green function and $N$ is the 
(infinite) number of points of the lattice.
(2) Next, we compute $\cal{G}(\omega)$ which includes self energy processes at 
all lattice sites except at the ``impurity'' site {\bf i} under 
consideration, 
${\cal{G}}^{-1}(\omega) = G^{-1}_{i,i}(\omega) + \Sigma_{i,i}(\omega)$.  
This step corresponds to a site-exclusion to prevent the over-counting of
self-energy diagrams on site $i$.  $\cal{G}(\omega)$ defines the undressed 
Green function of a generalized Anderson impurity model.
(3) We solve the associated impurity problem with some technique, e.g. 
the QMC-method, which produces $G_{imp}(\omega)$, the Green function of 
the generalized Anderson impurity model.
(4)  Then $\Sigma_{i,i}(\omega)= {\cal{G}}^{-1}(\omega) - G^{-1}_{imp}(\omega)$.
$\Sigma_{i,i}(\omega)$ may be used in (1) to continue the procedure.  The 
iteration typically continues until $G_{i,i}(\omega)=G_{imp}(\omega)$ to 
within the desired accuracy, and the procedure may be shown to be completely 
causal.

This DMFA algorithm may be applied in any dimension, but it is only exact for 
$D=\infty$. In finite dimensions, it is very difficult to formulate $1/D$ 
corrections to the DMFA which are both causal and systematic.  For example, 
consider the first non-trivial correction to the self energy of a Hubbard 
model on a hypercubic lattice given by the self energy diagrams evaluated 
between nearest neighbor sites $i$ and $j$. This contributes a term of 
order ${\cal{O}}(1/\sqrt{D})$ to the self energy which then assumes the form 
$\Sigma(\k,\omega)= \Sigma_{ii}(\omega) + \epsilon_\k\Sigma_{ij}(\omega)/t$, 
where $t$ is the hopping matrix element and $\epsilon_\k$ the bare electronic 
dispersion.  Note that when $\Sigma_{ij}(\omega)$ and/or $\epsilon_\k$ is 
large, it is possible for the imaginary part of the self energy ${\rm{Im}} 
\Sigma(\k,\omega)>0$, for some $(\omega, \k)$.  The corresponding 
quasiparticle excitations grow exponentially in time; a clear violation of 
causality.

\section {Dynamical Cluster Approximation}

For this reason, we formulated the DCA approach which includes systematic 
non-local corrections to the DMFA but is not systematic in $1/D$.  Like 
the DMFA, the DCA is a self-consistency scheme, although in the DCA the 
``impurity'' is replace by a finite-sized cluster.  Furthermore, the DCA 
is causal, and restores momentum conservation as well as the Ward 
identities systematically as the cluster size becomes large.

\begin{figure}
\epsfxsize=2.6in
\epsffile{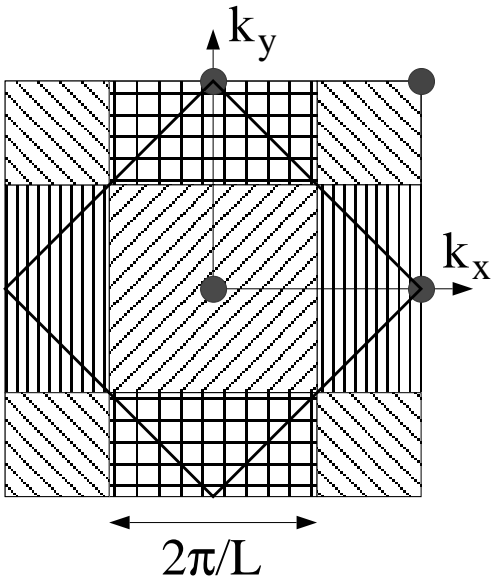}
\caption[a]{\em{The cluster momenta and coarse-graining cells for a 
$N_c=2\times 2$ cluster covering the Brillouin zone of a real two-dimensional
square lattice. The cluster momenta  are indicated by filled circles, and the cells
by different fill patterns.  The solid line in the shape of a diamond is the 
Fermi surface of the non-interacting system at half filling. 
The cells adjacent to the BZ 
boundary extend periodically to the opposite side.}}
\label{Kcluster}
\end{figure}

The general form of the DCA was given in Ref.~\cite{fkm}. Here, we briefly 
review the formalism, and then give a more detailed description of the method 
and its approximations.  For simplicity, we consider a single-band model with 
a local Hubbard-like interaction on a periodic hypercubic lattice with 
$N$ sites.  This is mapped onto a self-consistently embedded periodic
cluster of size $N_c=L^D$.  As illustrated in Fig.~\ref{Kcluster}, the 
corresponding crystal momenta $\K$ of the cluster are at the centers of a set 
of $N_c$ cells of size $(2\pi/L)^D$ inside the first Brillouin zone (BZ) for 
the lattice. Although there is considerable latitude in the choice of $\K$, 
we typically choose $K_{\alpha l}= \pi(2l/L-1)$ (where $l$ is an integer 
$1 \leq l \leq L$, and $\alpha$ indicates spatial direction)\cite{choiceK}.  

The crucial assumption of the DCA is that the irreducible self energy
of the cluster $\Sigma^c(\K,\omega)$ and the two--particle irreducible vertex 
functions of the cluster are good approximations to the irreducible self 
energy and vertex functions of the real lattice for values of the lattice
momenta inside the cells around the cluster momenta. This assumption is 
justified if the momentum dependence of the irreducible self energy 
and vertex functions of the real system is sufficiently weak; or 
equivalently, if the dynamical non--local correlations have a short range 
$b \alt L/2$. If this is the case, then, according to Nyquist's sampling 
theorem\cite{nyquist}, to reproduce these correlations in the self energy 
and vertex functions, we need only sample the reciprocal space at 
an interval of $\Delta k\approx 2\pi/L$; i.e., on a set of $N_c=L^D$ points 
within the first Brillouin zone.  Therefore, 
$\Sigma(\K+\tk,\omega)\approx \Sigma(\K,\omega)$ 
for each $\tk$ within a cell of size $\left( \pi/b\right)^D$ about $\K$, so
the lattice self energy is well approximated by the self energy 
$\Sigma^c(\K)$ obtained from the cluster. Similar arguments can be 
made for the vertex functions as well.
 
Next, within the spirit of the same approximation, the cluster self
energies and vertex functions can be equated with the {\em{coarse-grained
averages}} of the lattice self energies and vertex functions over these
momentum cells around the cluster momenta. For example, for the self energy,
\begin{equation}
\Sigma^c(\K,\omega) = \bar{\Sigma}(\K,\omega) =	
\frac{N_c}{N}\sum_{\tk}\Sigma(\K+\tk,\omega)
\label{sigmabar}
\end{equation}	
where the $\tk$ summation runs over the $N/N_c$ momenta of the cell 
about the cluster momentum $\K$.  This assumption is consistent with
that made in the previous paragraph, and insures that all the states
of the full system are represented once the problem is reduced to
the cluster.  Similar equations can be written down for 
the vertex functions.

The above two (related) sets of assumptions completely prescribe the DCA
and ensure that it reduces to an effective, self--consistently embedded 
cluster problem for any lattice problem with {\em{local}} interactions.  
For Hubbard-like models such as the HM, PAM and FKM, within a diagrammatic 
framework it is 
not hard to see that the skeleton graph expansions for the coarse--grained 
self energies and vertex functions defined above are then the same as the
skeleton graph expansions on a finite periodic cluster of size $N_c$.
The cluster Green function, $G^c(\K,\omega)$ is given by
the {\em{coarse-grained average}} of the Green function of the real lattice:
\begin{equation}
G^c(\K,\omega) = \bar{G}({\K},\omega) = \frac{N_c}{N}\sum_{\tk} \frac{1}
{\omega - \epsilon_{\K+\tk} +\mu - \Sigma^c(\K,\omega) } \,.
\label{gbar}
\end{equation}
Here, $\epsilon_\k$ is the dispersion for the noninteracting lattice problem
and $\mu$ is the chemical potential. The DCA assumption 
that $\Sigma(\K+\tk,\omega)\approx \Sigma^c(\K,\omega)$
has been explicitly put in for the lattice Green function.

One can now ask what bare Green function $\cal{G}(\K)$ on the cluster this
skeleton graph expansion corresponds to. The answer is determined by
the Dyson equation on the cluster used in reverse:
\begin{equation}
{\cal{G}}^{-1}(\K,\omega)= \bar{G}^{-1}(\K,\omega)+ \Sigma^c(\K,\omega).
\label{gsck}
\end{equation}
This step  corresponds to a ``cluster exclusion'' to prevent over-counting of 
self energy contributions from the interactions on the sites belonging to 
the cluster, analogous to the ``site exclusion''
of the DMFA ( which {\em{is}} the DCA if the cluster consists of a 
single site only). It is this step that determines the self--consistent
embedding of the cluster, since $\cal{G}$ includes the effects of
self energy processes at sites of the lattice other than the cluster sites,
and thus has strong retardation effects. The retardation effects can
be interpreted in terms of hybridization of the cluster (cells) to   
``conduction electron baths'' (one for each \K) analogously to the 
interpretation of the single site in DMFA in terms of an Anderson 
impurity problem.

\begin{figure}
\epsfxsize=3.5in
\epsffile{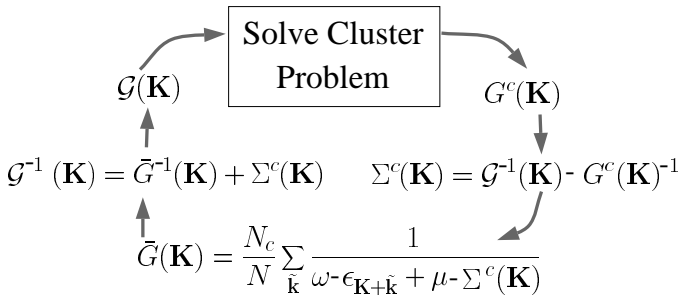}
\caption[a]{\em{Sketch of the DCA algorithm.}}
\label{DCA_alg}
\end{figure}

The DCA iteration procedure is now easily prescribed.  It is started by 
guessing an initial $\Sigma^c(\K,\omega)$, usually zero, which is used 
to calculate the coarse-grained Green function ${\bar{G}}(\K,\omega)$ using
Eq.~\ref{gbar}.  The cluster problem is then set up with the bare Green 
function $\cal{G}(\K,\omega)$ given by Eq.~\ref{gsck} and interactions on 
the cluster sites.  $\Sigma^c(\K,\omega)$ may then be calculated using any of 
a variety of methods, including perturbation theory, QMC, the non-crossing
approximation, etc., as appropriate. (If 
a skeletal graph perturbation expansion is used for the calculation, then the
cluster exclusion step may be skipped.)  For Green function techniques, such 
as QMC, which produce the fully-dressed cluster Green function 
$G^c(\K,\omega)$ 
rather than the self energy, the cluster self energy is calculated as
\begin{equation}
\Sigma^c(\K,\omega)= {\cal{G}}^{-1}(\K,\omega)- 
G^c (\K,\omega)^{-1}\,.
\label{sigc}
\end{equation}
The iteration closes by calculating a new $\bar{G}({\K},\omega)$ with 
Eq.~\ref{gbar},
and the iteration is continued until $\bar{G}(\K,\omega)=G^c(\K,\omega)$ to 
within the desired accuracy.  The self-consistency loop for the DCA is 
illustrated in Fig.~\ref{DCA_alg}.

In analogous fashion we can also provide prescriptions for calculating 
two--particle properties of the lattice from the irreducible cluster 
two--particle self energies (or vertex functions).
Again,  the basic assumption is that the momentum dependence of the
irreducible vertex function of the real lattice is weak. This 
is elaborated on in more detail in Appendix A.  

For lattice problems with non-local interactions 
such as the extended Hubbard Model, the problem is first
converted into one that has only local interactions by introducing
auxiliary Hubbard-Stratonovich Bosonic fields. The DCA can then
be prescribed in a straightforward way for this interacting
Fermionic-Bosonic problem with local interactions. The effective   
cluster problem will necessarily involve coarse-grained Bosonic Green
functions as well. The details are given in Appendix B.

\section{Discussion of the DCA}

In this section we provide a detailed discussion of some of the features 
of the DCA. We discuss the coarse-graining procedure and offer a simple 
diagrammatic interpretation. For large but finite $D$, we show that the DCA
includes short-ranged dynamical correlations without resorting to Nyquist's 
theorem, and we give a simple argument showing its causality.

\subsection{Coarse-Graining}
One can think of other (perhaps more ad-hoc) prescriptions for the
calculation of the cluster self energies and vertex functions, eg.,
using a modified $\bar{G}$ where the coarse-graining over $\k$ involves
a positive semidefinite weight function $f_w (\k, \K)$ which we can choose:
\begin{equation}
\bar{G}({\K},\omega) = \frac{1}{N}\sum_{\k} \frac{f_w (\k, \K)}
{\omega - \epsilon_{\k} +\mu - \Sigma^c(\K,\omega) },
\label{gbarprime}
\end{equation}
where the sum on $\k$ is now over the whole Brillouin zone.
Our choice  of 
\begin{equation}
f_w (\k, \K)=N_c\prod_l\Theta \left(\frac{\Delta k}{2}-|k_l-K_l|\right),
\end{equation}
where  $ \Delta k = 2\pi/L $
will reproduce the DMFA 
if the cluster is a single site. In addition,
even for larger clusters, the local lattice Green function and the local 
cluster Green function will be identical given our choice.  We note that 
the choice  $f_w (\k, \K)=N\delta(\k-\K)$ corresponds to evaluating
the system on the finite size cluster without any feed-back of the
host. For a cluster of one site this is identical to the atomic limit.
One could also imagine forms of $f_w$ which allow for overlap of the
cells in Brillouin zone, such as products of Gaussians.  However,
most $f_w (\k, \K)$ different from  the two specified above will lead to
a calculation which does not have an obvious physical limit for the 
case of a single site ``cluster''.

	The DCA also has a simple diagrammatic interpretation.
For Hubbard-like models, the local Hubbard $U$ is unchanged by the  coarse
graining, and thus the momentum dependence of each vertex is completely
characterized\cite{muller-hartmann} by the Laue function,
\begin{equation}
\Delta(\k_1,\k_2,\k_3,\k_4)=\sum_\r e^{i(\k_1-\k_2+\k_3-\k_4)\cdot\r},
\end{equation}
which expresses the conservation of momenta $\k_1$ and $\k_3$ ($\k_2$ and 
$\k_4$) entering (leaving) each vertex.
For example,
in the conventional diagrammatic approach $\Delta(\k_1,\k_2,\k_3,\k_4)
= N \delta_{\k_1+\k_3,\k_2+\k_4}$.  
If we reintroduce the cluster and cell momenta, such that $\k_i=\K_i+\tk_i$,
$i=1,4$, then 
\begin{eqnarray}
\Delta(\k_1,\k_2,\k_3,\k_4)&=&\sum_\r e^{i(\tk_1-\tk_2+\tk_3-\tk_4+\K_1-\K_2+\K_3-\K_4)\cdot\r}
\nonumber \\
      &=& N_c \sum_n \frac{1}{n!}\left((\tk_1-\tk_2+\tk_3-\tk_4)\cdot
      \grad_{K_1}\right)^n \nonumber \\
      && \;\;\;\;\;\; \times  \delta_{\K_1+\K_3,\K_2+\K_4}.
\end{eqnarray}
Within the DCA, only the first term in the sum ($n=0$)is kept so 
\begin{eqnarray}
\Delta_{DCA}(\k_1,\k_2,\k_3,\k_4)&=& 
N_c \delta_{\M(\k_1)+\M(\k_3),\M(\k_2)+\M(\k_4)}\nonumber \\
&=& \Delta(\k_1,\k_2,\k_3,\k_4)+{\cal{O}}(\Delta k),
\label{laue_DCA}
\end{eqnarray}
where $\M(\k)$ is a function which maps $\k$ onto the momenta label $\K$
of the cell containing $\k$.  Note that with this choice of Laue function 
the momenta of each internal leg may be freely summed over the cell.  Thus, 
each internal leg $G(\k_1,\omega)$ in the diagram is replaced by 
$ \bar{G}(\M(\k_1),\omega) $ defined by Eq.~\ref{gbar}. 
Furthermore, since each external momenta $\k$ also enters the diagram only
through $\M(\k)$, the self energy becomes momentum independent within 
each cell, {\em{i.e.}} it obtains the coarse-grained form defined in
Eq.~\ref{sigmabar} and the approximation $\Sigma(\k,\omega)\approx
{\bar{\Sigma}}(\M(\k),\omega)$ follows as a natural consequence.  
In the DMFA, the cell momenta extend over the entire Brillouin zone,
so that $\Delta_{DMFA}(\k_1,\k_2,\k_3,\k_4)=1$\cite{muller-hartmann}
and momentum conservation is neglected.
Thus, the above choices of the Laue function serve as  
microscopic definitions of 
the DCA, and of the DMFA.  To interpret the choice for the DCA,
note that small changes in each of the 
internal momentum labels will not affect $\Delta_{DCA}$.  Thus, momentum 
conservation for small momentum transfers less than $ \Delta k=2\pi/N_c^{1/D}$ 
is neglected.  However, note that for momentum transfers larger than
$\Delta k$ momentum conservation is (partially) observed at the
vertex. Thus, the DCA systematically restores the momentum 
conservation relinquished by the DMFA as the cluster size increases.

\subsection{Non-local corrections}
The range of the dynamical correlations included in the DCA is
dictated by the cluster size and by the
range of the Green functions used to calculate the irreducible graphs.
In the DMFA, the self energy is a functional of the local Green function,
but in the DCA non-local Green functions also are used.  Thus, the DMFA
incorporates only local dynamical correlations which occur on the
effective impurity, whereas the DCA incorporates non-local
dynamical correlations which occur on the cluster.  

This may be seen by exploring the coarse-graining step in detail, and in 
real space.  For this purpose, we consider a lattice in large but finite 
$D$ which we divide into $L^D$-sized clusters.
Let $\r$ denote vectors within a cluster, and $\R$ the vectors between
the centers of the clusters. The points of the original lattice
can be represented as ${\R}+{\r}$.
The relation between the real Green function $G({\R}+{\r},\omega)$
and the cluster Green function $\bar{G}(\r,\omega)$ is given by
\begin{equation}
\bar{G}(\r,\omega) = \frac{1}{N} \sum_{\K,\tk} \sum_{\R,\r'} e^{i \K \cdot (\r-\r')}
e^{-i \tk \cdot (\R+\r')} G({\R+\r'},\omega) \,.
\end{equation}
The sum over $\K$ forces $\r'= \r$.  For $\R=0$ the additional phase 
factor $e^{-i\tk\cdot\r}$ is essentially 1 over the entire range of $\tk$
for short distances on the cluster  $r \ll 2\pi/\Delta k$, which
leads to a contribution to $\bar{G}(\r,\omega)\approx G(\r,\omega)$.  
Contributions from larger $\R$ are suppressed both by the oscillations in the
phase factor which suppresses the integral and from the smallness of 
$G(\R+\r',\omega)$ itself. More precisely, with the choice 
$K_{\alpha l}= \pi(2l/L-1)$ 
(where $l$ is an integer $1<l<L$, and $\alpha$ indicates spatial direction),
we can complete the sums on momenta exactly to obtain
\begin{equation}
\bar{G}(\r,\omega)=\sum_{\R} \prod_{l=1}^{D}\left(\frac{\sin[\pi (x_l+X_l)/L]}
{\pi (x_l+X_l)/L}\right) G(\R+\r,\omega) \,,
\end{equation}
where $x_l$ ($X_l$) is the $l$-th component of the vector $\r$ ($\R$). 
Thus, $\bar{G}(\r,\omega)$ is composed of a sum over $G(\r+\R,\omega)$ with 
each term weighted by a sinusoidal prefactor that falls off like 
${\mid \r+\R \mid}^{-D}$.  For small $r$, the leading term in the sum comes 
from $\R=0$.  Then, by expanding the sinusoidal prefactor, we can see 
that for $\r=0$, $\bar{G}({\bf{0}},\omega)=G({\bf{0}},\omega)$, and for 
$r\ll L/2$, 
$\bar{G}(\r,\omega)\approx G(\r,\omega) +{\cal{O}} \left( (r \Delta k)^2 \right)$. 
Contributions from $G(\r+\R,\omega)$ for finite values of $\R$
are cut-off by the sinusoidal prefactor and the exponential fall-off of 
the Green function itself, since  for large distances $G(r)\sim D^{-r/2}$.  
Thus, short-ranged correlations are accurately represented by 
$\bar{G}(\r,\omega)$, and longer ranged contributions are cut off.

\begin{figure}
\epsfxsize=2.6in
\epsffile{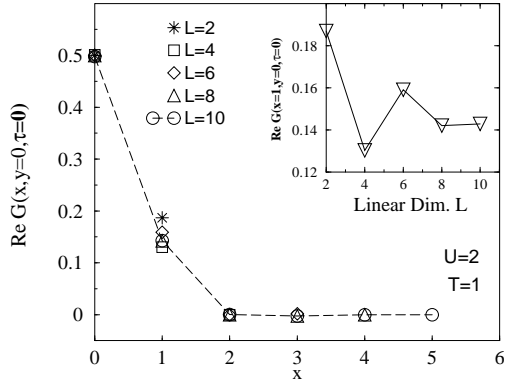}
\caption{${{\rm{Re}}\, \bar{G}}(x,y=0,\tau=0)$ versus $x$ for various 
cluster sizes, obtained from QMC simulations of the FKM.  In the inset, 
${\rm{Re}}\, \bar{G}(x=1,y=0,\tau=0)$ is plotted versus cluster size 
(linear dimension $L$). } 
\label{Gbarofr}
\end{figure}

	This behavior is seen even in two-dimensional systems, a shown in 
Fig.~\ref{Gbarofr} where $\bar{G}(x,y=0,\tau=0)$ calculated with a QMC 
simulation of the two-dimensional half-filled FKM (see Sec.~V) is plotted 
versus $x$ for various cluster sizes.
The $\r=0$ result is fixed by the filling, $\bar{G}(x=0,y=0,\tau=0)=0.5$;
however the near neighbor result shows some significant dependence on
the cluster size.  $\bar{G}(x=1,y=0,\tau=0)$ is plotted versus the linear
cluster size in the inset to Fig.~\ref{Gbarofr}.  Note that it quickly 
converges to $\bar{G}(x=1,y=0,\tau=0)\approx0.143$ as the cluster size 
increases, indicating that short-ranged correlations are correctly described 
by the DCA for this model.  For larger $x$, $\bar{G}(x,y=0,\tau=0)$ falls
quickly to nearly  zero.
	
\subsection{The Role of Reducible and Irreducible Quantities}

	In Appendix D, we show that the DCA (and the DMFA)
is not conserving, thus the calculations of different measurable quantities
are not unique.  For example, we approximate the lattice self energy
$\Sigma(\k,\omega)\approx {\bar{\Sigma}}(\M(\k),\omega)$, and calculate the 
Green function using 
$1/G(\k,\omega)=1/G^0(\k,\omega)-{\bar{\Sigma}}(\M(\k),\omega)$; however,
a different approximation, corresponding to a different implicit
choice for $\Sigma(\k,\omega)$ would be to approximate
$G(\k,\omega)\approx \bar{G}(\k,\omega)$.  We show in
Appendix A, that the former prescription is the unique choice which minimizes
the DCA free energy, and thus is the correct choice.
A similar problem exists for the calculation of two--particle
properties such as the magnetic susceptibility.  However, as discussed
in Appendix A, the approximation 
$\bgam\approx \bar{\bgam} \equiv \delta\bar{\bSig}/\delta\G$
for the lattice two--particle vertex yields an estimate for the susceptibility 
(Eq.~ A15) equivalent to
that calculated from the second derivative of the free energy with
respect to the external field.

Thus, in general, {\em{the cluster calculation should only be used to 
provide the irreducible quantities}}.  These, together with the 
bare real-lattice Green functions, may be used to construct the corresponding 
reducible quantities.

At least for the single-particle Green functions, this prescription may also 
be motivated physically.  Short ranged correlations are accurately represented 
by the cluster irreducible single-particle self energy.  Following the 
discussion of the last section, one may show that for $r\ll L/2$,
$\Sigma^c(\r,\omega)\approx 
\Sigma(\r,\omega) + {\cal{O}}\left((r\Delta k)^2\right)$, since it is 
calculated from cluster quantities.  In addition, since 
the self energy is formed from higher order products of the Green function, 
e.g. $\Sigma(\r)\sim \left[G(\r)\right]^3\sim D^{-3r/2}$ for the second order
contribution in the Hubbard model,
in high dimensions it falls faster with increasing $r$ than the Green function 
itself.  Thus, the correction terms coming from $\R\neq 0$ will be smaller 
for irreducible quantities such as the self energy than it will be for
reducible quantities like the Green function.   
Since the range of the correlations that are treated increases with the 
cluster size, away from a transition, the irreducible quantities calculated 
on the cluster will have converged to acceptable values before their 
reducible counterparts.

	Finally, we note that while in the last two subsections we used
$1/D$ arguments to justify the approximations made in the DCA, the DCA is 
not systematic in $1/D$.  For example, even for short distances $r$, which 
would correspond to low orders in $1/D$, $\bar{G}(\r,\omega)$ contains 
contributions $G(\r+\R,\omega)$ corresponding to much larger distances and 
higher orders in $1/D$.  Furthermore, since the density of states of the 
finite-dimensional lattice is used to calculate the host propagator 
${\cal{G}}$, the approximation includes corrections to all orders in $1/D$.  
In fact, we have shown in this subsection that the cluster quantities 
differ from those of the real lattice by terms of order 
$(\Delta k)^2 = 4\pi^2/N_c^{2/D}$.  Thus, the DCA is a systematic approximation
in $1/N_c$, not $1/D$.

\subsection{Causality}
We can also show that the DCA algorithm is fully causal, i.e. that the spectral
weight is conserved and that the imaginary parts of the single-particle 
retarded Green
functions and self energies are negative definite.
 Here, since many methods can be
used to solve the cluster problem, we will assume that all are causal, i.e.,
given a causal ${\cal{G}}$, then the resulting $\Sigma^c$ and $G^c$ are
also ensured to be causal by the method chosen to solve the cluster problem.
Furthermore, ${\bar{G}}(\K,\omega)$ is causal since $\Sigma^c(\K,\omega)$ is causal.  
Thus,
Eq.~\ref{gsck} is the only step in the algorithm where problems with causality 
could occur.   In Ref.~\cite{fkm} we argued using a continued fraction
expansion that the $\tk$ averaging (coarse--graining)
of Eq.~\ref{gbar} adds a causal piece to the self energy of $\bar{G}$ 
that allows $\cal{G}$ to remain causal even after the subtraction of 
$-\Sigma^c(\K,\omega)$ in Eq.~\ref{gsck}. Here, we give a simple {\it 
geometrical} argument (which is recast as a formal proof in Appendix C) that 
causality holds for rather general models, including the HM and the FKM.

\begin{figure}[htb]
\epsfxsize=2.6in
\epsffile{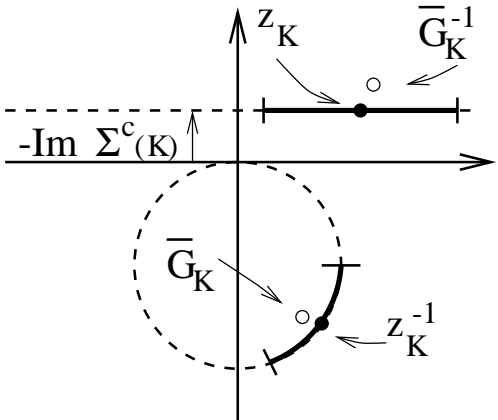}

\caption{Illustration of the essential steps of the proof that the DCA is
causal (see text).}
\label{causal}
\end{figure}

There are two steps to the argument: first, we must show that weight 
is conserved,
and second, that the imaginary part of ${\cal{G}}$ is negative semidefinite.
The first part follows from the causality of $\Sigma^c$ and $\bar{G}$ which
both fall off inversely with frequency at large $\omega$, and
in particular $\bar{G} \sim 1/\omega$.  From Eq.~\ref{gsck} it is then
apparent that ${\cal{G}} \sim 1/\omega$ so that spectral weight is preserved.  
The second part of the argument is sketched in Fig.~\ref{causal}.  The 
imaginary part of 
${\cal{G}}(\K,\omega)=(\bar{G}(\K,\omega)^{-1} + \Sigma^c(\K,\omega))^{-1}$ is 
negative provided that 
${\rm Im} (\bar{G}(\K,\omega)^{-1}) \geq -{\rm Im}\Sigma^c(\K,\omega)$. 
$\bar{G}(\K,\omega)$ can be written as
$\bar{G}(\K,\omega)=(N_c/N) \sum_{\tk} (z_{\K+\tk})^{-1}(\omega)$, where
the $z_{\K+\tk}(\omega)$ 
are complex numbers with a positive semidefinite imaginary part 
$-\mbox{Im}\Sigma^c(\K,\omega)$.
For any $\K$ and $\omega$, the set of points 
$z_{\K+\tk}(\omega)$ are on a segment of  the 
dashed {\it horizontal} line  in the upper half plane 
due to the fact that the imaginary part is {\it independent} of $\tk$.
The mapping $z\rightarrow 1/z$ maps this line segment onto a segment
of the dashed circle shown in the lower half plane. 
$\bar{G}(\K,\omega)$ is obtained by summing the points on the
circle segment, 
yielding the empty dot that must lie {\it within} the dashed circle.
The inverse necessary to take $\bar{G}(\K,\omega)$ 
to $1/\bar{G}(\K,\omega)$ maps this
point onto the empty dot in the upper half plane which must lie {\it above}
the dashed line.  
Thus, the imaginary part of $\bar{G}(\K,\omega)^{-1}$ is greater than or 
equal to $-{\rm Im}\Sigma^c(\K,\omega)$.  This argument may easily be extended
for ${{\cal{G}}(z)}$ for any $z$  in the upper half plane. Thus ${\cal{G}}$
is completely analytic in the upper half plane.

\section{DCA for the Falicov-Kimball model.}

	Here, we illustrate the power of the DCA with a QMC simulation of 
the two-dimensional Falicov-Kimball model.  The FKM is studied, instead of,
for example, the much more complicated Hubbard model
(for which there is work in progress\cite{hossein}), for several reasons.
First, the FKM is perhaps the simplest model of correlated electrons which 
retains a complex phase diagram, including a Mott transition and a charge
density wave (CDW) ordering transition\cite{deraedt}.  Second, it has 
been extensively studied by de Vries {\em{et al.}} with QMC 
simulations\cite{deraedt_linden}
of finite-sized systems which may be compared to our results.  Third, it 
is possible\cite{deraedt_linden} to calculate the real-frequency spectra 
without the need for computationally expensive numerical analytic 
continuation. Finally, it is of considerable experimental 
interest\cite{NiI2_expt}.

The FKM can be considered as a simplified Hubbard model in which one spin 
species is prohibited to hop.  In the particle--hole symmetric case the 
Hamiltonian reads
\begin{equation}
H = - t \sum_{<i,j>} d^{\dagger}_i d_j -\mu \sum_i 
(n^d_i +n^f_i) + 
U  \sum_i n^d_i n^f_i \ , 
\end{equation} 
with $n^d_i = d^{\dagger}_i d_i$, $n^f_i = f^{\dagger}_i f_i$, and  $\mu=U/2$.
For a 2D square lattice with nearest neighbor hopping ($\langle i,j \rangle$)
the dispersion is $\epsilon_\k= -2 t (\cos{k_x} +\cos{k_y})$. We measure 
energies in units of the hopping element $t$.
Consequently, the bandwidth of the noninteracting system is $W=8$.  For 
$D \ge 2$ the system has a phase transition from a homogeneous high 
temperature phase with  
$\langle n^d_i\rangle =\langle n^f_i\rangle =1/2$ to a checkerboard phase 
(a charge density wave with ordering vector ${\bf{Q}}=(\pi,\pi,...)$) with  
$\langle  n^d_i\rangle \ne \langle n^f_i\rangle$ for $0 < U < \infty$. 
\cite{brandt1}

\subsection{Exact Enumeration} 

In contrast to the Hubbard and related models, the DCA for the
FKM can be solved without the application of QMC since the f-electrons are 
static, acting as a kind of annealed disorder potential to the dynamic d-electrons.  
Here, we generalize the algorithm of Brandt and Mielsch \cite{brami} to a 
finite size cluster. We first compute the Boltzmann weights
$w_f$ of all configurations $\{f\}$ 
of f-electrons on the cluster, given an initial host
Green function ${\cal{G}}_{ij}$ of the d-electrons via $w_f=w_f^0/Z$ where
\begin{eqnarray}
w_f^0= 2^{Nc}\prod_{\omega_n} \det \frac{{\cal{G}}^{-1}_{ij}(i\omega_n) - U n^f_i
\delta_{ij}}{i\omega_n \delta_{ij}}
\end{eqnarray}
is the unnormalized weight, and $Z= \sum_{\{f\}} w_f^0$ is the 
``partition sum''. The determinant is to be taken over the spatial indices.
This expression is written such that the 
product converges at large frequencies. Given the weights, the new
d-electron cluster Green function is given by
\begin{equation}
G^c_{ij}(z) = \sum_{\{f\}} w_f \left[{\cal{G}}^{-1}_{ij} (z) - U n_i^f
\delta_{ij} \right]^{-1}
\end{equation}
for an arbitrary complex frequency argument $z$, in particular also for
$z=i\omega_n$ (Matsubara) and $z=\omega +i\eta$ (retarded). 
The self--consistency loop closes by use of the Eqs. \ref{gbar}
,\ref{gsck} and \ref{sigc}. 

Because the number of f-configurations grows exponentially with the 
cluster size the exact enumeration method is  confined to
small clusters (up to $4\times4$ in the broken symmetry state, see below).  
We first simultaneously determine the  weights and the  Matsubara Green 
function. Then, we use knowledge of the weights to find the retarded Green 
function. Convergence of the algorithm is fast for Matsubara frequencies, but 
relatively slow for real frequencies.

\subsection{Quantum Monte Carlo} 

The FKM is particularly suitable to a QMC evaluation of the configuration 
sums since the f-electrons are themselves like classical Ising spin 
variables. Following De Raedt and von der Linden\cite{deraedt_linden}, 
given a particular configuration, we can propose ``spin flips'', 
corresponding to a change of the f-occupation $n^f_i\to 1-n^f_i$ at a single 
site $i$. The ratio $R$ of weights $w_f'$ of the proposed configuration to 
the weight $w_f$ of the original configuration  is  (at half filling)
\begin{equation}
R = \prod_{\omega_n > 0} (1 - \lambda_i G^c_{i,i}(i\omega_n))
(1 - \lambda_i G^{c*}_{i,i}(i\omega_n)),
\label{ratio}
\end{equation}
with $\lambda_i= - U s(i)$ and  
$s(i) = 2 n^f_i - 1$.  Note that the ratio $R$ is always real and positive 
since the Matsubara Green function is Hermitian $G^c_{i,i}(-i\omega_n)=
G^{c*}_{i,i}(i\omega_n)$. This holds for any filling. Consequently, there 
is no sign problem as there is, e.g., in the Hubbard model away from 
half filling. 

A configuration change is accepted by comparing a random number in the 
interval $(0,1)$ to $R/(1+R)$ (``heat bath method'') or to $R$ itself 
(``Metropolis method''). Once the change at site  $i$ is accepted, the 
Green function is updated via
\begin{equation}
 G'^c_{j,k}(i\omega_n) =  G^c_{j,k}(i\omega_n) + 
 \frac{\lambda_i G^c_{j,i}(i\omega_n)
\otimes G^c_{i,k}(i\omega_n)}{1 - \lambda_i G^c_{i,i}(i\omega_n)},
\label{update}
\end{equation}
where $\otimes$ denotes a direct matrix product (no summation).
Most of the total CPU time is consumed by this updating step. However,
the fact that we can work with  frequencies rather than imaginary time
drastically reduces the amount of time required.  Note that although 
Eq.~\ref{update} is written for Matsubara Green functions
an analogous relation holds for the real frequency Green functions
which allows us to calculate dynamical properties without the
need for analytic continuation.  On the other hand the ratio $R$ is 
completely determined by the Matsubara Green function. This means that 
we determine the acceptance from the Matsubara Green function and 
then update both the Matsubara and the real-frequency retarded
Green function ``simultaneously''.

The measurement of the two--particle properties consumes large 
amounts of memory and CPU time. Since they are not required for the
self consistency cycle (Fig.~2), they are measured only after convergence 
of the single-particle properties. In fact, due to the enormous size 
of the susceptibility matrix it is often worthwhile to separate the single--
and two--particle calculations to different computer runs. 

\section{Results}

\begin{figure}
\epsfxsize=2.6in
\epsffile{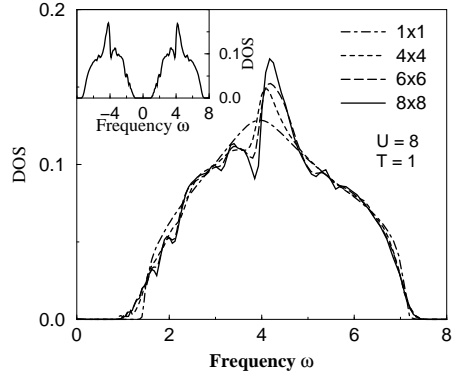}
\caption{ Local density of states for various cluster sizes.
The density of states is essentially converged for the $6\times6$--cluster,
though some fine structure near $\omega=\pm U/2$ continues to emerge for 
the larger cluster sizes (see discussion in text).
}
\label{localdos}
\end{figure}

	In this section, we present results from both exact enumeration and 
QMC simulation of the two-dimensional FKM for a variety of parameters and 
cluster sizes.  There is considerable latitude in the selection of the 
cluster momenta. This is because i) the sites on the cluster do not really 
correspond to the physical lattice, and ii) because for large clusters any  
differences due to this choice should vanish.  Here, for an $L\times L$ 
cluster we choose either $K_{\alpha l}= \pi(2l/L-1)$, or 
$K_{\alpha l}= \pi(2l/L-1)-\pi/L$ (where $l$ is an integer $1\leq l\leq L$,
 and 
$\alpha=x\;{\rm{or}}\; y$).  These choices, respectively, correspond to 
periodic or antiperiodic boundary conditions for the cluster Green function 
$G^c(x+L,y,\omega)=\pm G^c(x,y,\omega)$. We use antiperiodic boundary 
conditions only for some data in Fig. \ref{tcplot}.

\subsection{Density of states and spectral function}
\begin{figure}
\epsfxsize=2.6in
\epsffile{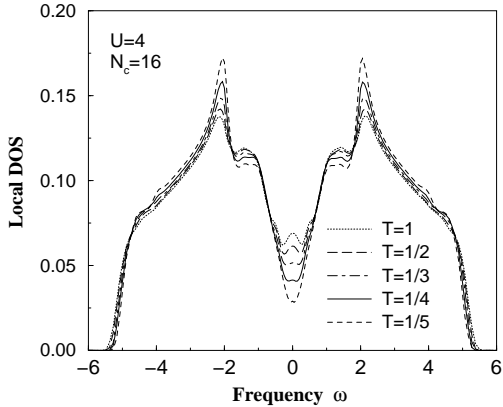}
\caption{ Local density of states when $U=4$ for a $4\times4$--cluster
at various temperatures. The DOS develops a pseudogap as the temperature 
approaches $T_c \approx 0.189$. This shows the influence of the non--local
CDW fluctuations present in the DCA $(N_c >1)$. In the DMFA 
($N_c =1$), there is no $T$--dependence of the DOS above $T_c$.
}
\label{dosvst}
\end{figure}

\begin{figure}
\epsfxsize=2.8in
\epsffile{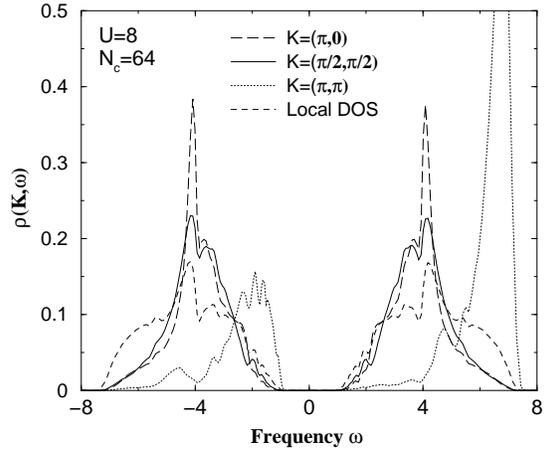}
\caption{Spectral function $\rho(\K,\omega)$ for various cluster momenta 
$\K$. 
Note the three
peak feature for $\K=(\pi,\pi)$ at the upper edge of the lower band. 
}
\label{dosk}
\end{figure}

We begin by discussing the (local) density of states (DOS) and the
$\K$--dependent spectral function shown in Figs. \ref{localdos}--\ref{dosk}.
In Fig. \ref{localdos} we show the local DOS for various cluster sizes up 
to $8\times8$ for the half-filled model and display only the positive 
frequencies.  The full spectrum is symmetric, due to particle-hole
symmetry, as shown in the inset.  With the exception of a peak which
develops at $\omega=\pm U/2$, the spectrum converges quickly as $N_c$
increases.  In fact, the convergence to the thermodynamic 
limit is apparently much faster than that seen in finite-sized lattice 
simulations\cite{deraedt}, where even for an $8\;\times\; 8$ system, the 
broadened spectra are often composed of a set of discrete spikes.

Furthermore, the DOS develops three distinct primary features also seen
in the finite-size calculations\cite{deraedt}.  First, 
as shown in Fig.~\ref{localdos}, for large $U\agt U_M$ the DOS develops a 
Mott gap centered at $\omega=0$, even though $T \gg T_c$.  The value of 
$U_M$ at this temperature changes slowly with cluster size, with 
$U_M\approx 5$.  Second, as shown in Fig.~\ref{dosvst}, for $U<U_M$, upon 
decreasing the  temperature, the DOS for $N_c>1$ develops a pseudogap at 
the Fermi energy associated with charge ordering fluctuations.  This 
pseudogap is absent when $N_c=1$ (as are the charge ordering fluctuations),
and it becomes more pronounced as the cluster size increases.
Third, as the charge ordering becomes more pronounced, either
by lowering the temperature or increasing the cluster size,
a sharp peak begins to develop in the DOS shown in Figs.~\ref{localdos}
and \ref{dosvst} at $\omega=\pm U/2$. In the ordered state, each
occupied f (d) orbital is surrounded by four occupied d (f) orbitals.
Thus, for large $U$ and low $T$ the electrons become highly localized
so the spectrum will develop very narrow ``atomic'' peaks at $\omega=\pm U/2$.

In addition, there are a surprising number of smaller features which emerge 
in the DOS. This is true even for the largest cluster, in some sense even 
more so, as some fine structure in Fig.~\ref{localdos} seems to develop for  
the $8\times8$--cluster that was only vaguely present for smaller clusters. 
This fine structure is more visible in the momentum--resolved spectral
function $\rho(\K,\omega)=\frac{1}{\pi} {\rm{Im}} G(\K,\omega)$, see 
Fig.~\ref{dosk}. In particular, note the 
three peak feature at negative frequencies for $\K=(\pi,\pi)$. Of course,
we really don't know how the DOS for the infinite lattice is supposed 
to look like. The extremely smooth form
the DMFA  provides is  mostly due to the lack of 
associated energy scales. In the DCA, we have at least $U$ and $J=t^2/2U$, 
and, in principle, many other scales can be constructed representing 
collective excitations of the cluster charges. That such features
emerge as the cluster size is increased can be understood by the following
argument. In addition to the self energy arising from interactions on the 
cluster the host also provides a self energy and therefore a broadening.
Consequently, features that are in principle present for smaller 
clusters like $4\times4$
are washed out by the host's broadening. Only as the host becomes  
less important (as cluster size increases) do the smaller energy features 
emerge from the background.

\subsection{Phase diagram and finite size scaling}
We now discuss the phase diagram and its dependence on cluster size.
In Ref. \cite{fkm} we showed that the transition temperature 
of the CDW--transition was significantly suppressed with respect to the DMFA 
when non--local correlations come into play. We have since extended this
analysis in two directions. 

In Ref. \cite{fkm} the result for the
$2\times2$--cluster was computed via the exact enumeration method
in the broken symmetry phase. This means we actually simulated two 
$2\times2$--clusters forming a bipartite cluster of $2\times2\times2=8$ sites.
The extension of the above described exact enumeration  method is 
straightforward and involves Green functions that are now $2\times2$ matrices
with respect to the bipartite cluster  (A and B sublattice index).
$T_c$ was then obtained by three steps: 1) We apply a staggered field 
at low enough temperatures (below the expected $T_c$) to drive the system into
the broken symmetry state with 
$\langle n^d_{i \in A} \rangle \neq \langle n^d_{j \in B} \rangle$ . 
2) We remove the staggered field. The system
relaxes but stays in the broken  if $T < T_c$. 3) We increase $T$ until
the system enters the uniform phase with 
$\langle n^d_{i \in A} \rangle = \langle n^d_{j \in B} \rangle$.
This method is very precise, but for larger clusters very time consuming.
Using the QMC method in the broken symmetry phase is possible, but 
 $T_c$ can not be determined precisely due
to critical fluctuations.
So the above described method is limited to at most $4\times4$--clusters, or
a total of 32 sites. This also means that a systematic finite size analysis
with this method alone would not be possible.

In order to get $T_c$ for larger clusters
we choose a different route. We compute the 
staggered charge  susceptibility $\chi(\Q=(\pi,\pi))$ with the method 
discussed in Appendix A.  Because the host always provides a mean field 
environment, the susceptibility diverges as
$\chi(\Q=(\pi,\pi)) \propto (T -T_c)^{-\gamma}$ with a mean field exponent 
$\gamma=1$ for $T$ close enough to $T_c$.
(Critical fluctuations cause $\gamma$ 
to deviate from the mean field value for somewhat larger values of $T-T_c$.)
This again allows a precise estimate of $T_c$.   The computational drawback 
here is the enormous memory requirements of the susceptibility matrix
needed at intermediate steps of the calculation.

\begin{figure}
\epsfxsize=2.6in
\epsffile{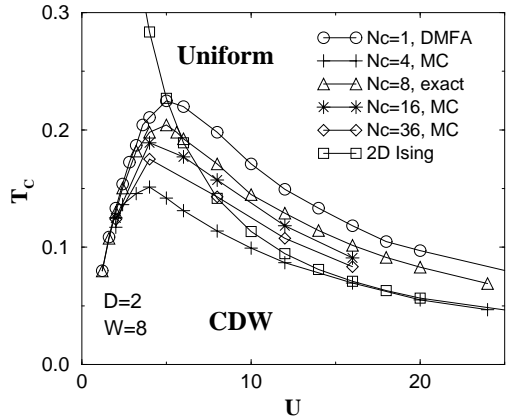}
\caption{ Phase diagram for various cluster sizes $N_c$. 
With the exception of $N_c=4$ (see text) the $T_c$ monotonically converge
with increasing cluster size. At large $U$ the system
maps to a 2D Ising model with $J=1/(2U)$. 
}
\label{phase}
\end{figure}

\begin{figure}
\epsfxsize=2.6in
\epsffile{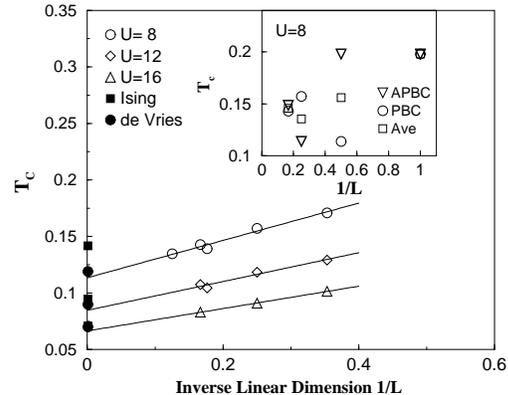}
\caption[a]{{$T_c$ as a function of inverse linear cluster dimension for 
the larger clusters and various $U$.  The Ising limit, and de Vries 
{\em{et al.}} \cite{deraedt} estimates of $T_c$ from 
simulations of finite-sized clusters are shown for comparison.  The 
extrapolated $T_c$'s generally fall below the finite-size estimates as well 
as the Ising limit (which should serve as an upper bound and become exact for 
large $U$).
The inset shows the influence of the cluster boundary conditions on $T_c$. 
The effect of boundary conditions becomes smaller with increasing 
cluster size.
}}
\label{tcplot}
\end{figure}

After these preliminaries we now discuss the results of these calculations 
in Fig. \ref{phase} and Fig. \ref{tcplot}.  Fig. \ref{phase} shows the 
phase diagram for various cluster sizes, all of them equipped with periodic
boundary conditions (PBC). In addition, we show the $T_c$ of the
2D Ising model given by $T_c^{Ising} =2.268 J$ with 
a coupling $J= 1/(2U)$.
We show the Ising result because the half filled FKM reduces to an Ising
model with such a coupling in the limit of large $U \gg W$. 
The FKM data are all 
obtained from the evaluation of the susceptibility with the MC--method
except for the $N_c=8$
data which are obtained by the exact enumeration method in the broken 
symmetry (two $2\times2$--clusters). For the DMFA the two methods give 
identical results (within 1\% accuracy). The phase boundary has
always the same general shape for the FKM data, with a slightly cluster
size dependent maximum at about half the bandwidth $W$. 

The results from
the MC--method converge monotonically with cluster size with one notable 
exception: The $2\times2$--cluster ($N_c=4$) has the lowest $T_c$ of all,
and even seems to fall below the Ising results for all $U$.
The reason for this exceptional behavior are not entirely clear to us.
At first one might consider a double counting of neighbors and a resulting
doubling of the energy scale common in standard lattice methods to be 
the reason. But clearly, the $T_c$'s of all clusters agree well at 
small $U$ where only local correlations are important. 
This rules out a simple doubling of the energy scale. A likely reason
for this unusual behavior lies in the particular 
way the BZ is sampled in the $2\times2$--cluster, see Fig.~\ref{Kcluster}. 
The only points on the Fermi surface are  $\K=(\pi,0), (0,\pi)$. These, 
however, are also the points responsible for the van Hove--singularity of 
the noninteracting system. In comparison to other momenta on the Fermi 
surface these points have extraordinary large scattering rates, making
them unfavorable for the formation of CDW--fluctuations driving the 
transition.  As a consequence, the  $T_c$ for this cluster is exceptionally low.

Although the $T_c$ results from a given method are monotonically
decreasing (with the one exception noted above) it is not obvious
how to scale the data as a function of cluster size; for, to
our knowledge, a rigorous
finite size scaling theory for a {\em {quantum-dynamical}} cluster
coupled to a {\em {quantum-dynamical}} host does not exist. 
However, such questions have been addressed in the context of 
systematic self--consistent cluster approximations for 
{\em {classical}} statistical systems, in particular, the  2D
Ising model \cite{suz} , which should be relevant to our problem,
at least for large U. Furthermore, on general grounds one expects
that for critical phenomena at {\em {finite temperatures}} the asymptotic 
scaling properties even of a {\em {quantum}} system
will be determined by the same universality class as for 
the corresponding {\em {classical}} system 
(i.e., with the same order parameter symmetry and the same spatial 
dimensionality). Hence, one expects \cite{suz} that our results for
$T_c (L) - T_c (\infty)$ should scale asymptotically
as $L^{-1/\nu}$, i.e., as $1/L$, since $\nu =1$ for the $2D$ Ising Model.
In Fig. \ref{tcplot} we therefore plot the $T_c$ data as a 
function of $1/L$ (or $1/\sqrt{N_c}$ for the broken symmetry results).
In the main part of the plot we show the results for large clusters with 
PBC which scale approximately linearly with $1/L$. 
The $N_c=32$ result (broken symmetry)
for $U=8$ and $U=12$ is a bit lower than the $T_c$ for $N_c=36$ (MC).
This shows that the two methods are not easy to combine, 
but the difference seems
small enough not to disrupt the predominant linear scaling with $1/L$.

For $U=16$ the cluster $T_c$'s scale well and the extrapolation to the 
infinite system comes very close to the Ising limit (or the results of 
de Vries {\em{et al.}}).  For smaller $U$ the Ising model is not 
appropriate, and it shows, as the Ising $T_c$ is much higher than the 
extrapolated $T_c$ of the clusters.  However, the extrapolated cluster
results are very close to the results obtained from finite-sized lattice
simulations.  The fact that the cluster estimates of $T_c$ consistently 
fall below de Vries results is likely due to finite-sized effects (
de Vries {\em{et al}} simulated lattices of up to 64 sites).  We also note 
that the $T_c$'s of the  $2\times2$--cluster
(not shown in Fig. \ref{tcplot} are in excellent agreement with
the cluster extrapolated values and the Ising result for large $U$. 
We have currently no explanation for this 
phenomenon. Though probably pure coincidence, the
fact remains:  the $T_c$ of the $2\times2$--cluster seems to provide
a good estimate of the $T_c$ of the $D=2$ FKM.

The inset shows the same $T_c$'s as in the main plot (all determined via MC) 
for $U=8$ of various cluster sizes,  and in addition the $T_c$'s for the
same clusters equipped with antiperiodic boundary conditions (APBC). As noted
before, the DCA does not intrinsically determine the choice of cluster
momenta. But different choice of cluster momenta will also in general
affect $T_c$ and other quantities. As PBC and APBC seem to span the entire
range it is interesting to see by how much the $T_c$'s differ. As illustrated
in the inset it matters quite a bit for very small clusters, but not
much once we consider clusters of the $6\times6$ size\cite{average}. 
The difference for
$2\times2$--clusters is extreme for the following reason: we noted above
that the $2\times2$--cluster with PBC has the lowest $T_c$ of all clusters 
with PBC. The $2\times2$--cluster with APBC, on the other hand, is 
by symmetry identical to the single site cluster which has the maximum 
$T_c$. Similarly, the $4\times4$--cluster with APBC
is by symmetry identical to the $2\times2$--cluster with PBC. But
once we go to cluster sizes  beyond this such identifications are
no longer possible. Concurrently, the $T_c$'s of the clusters also 
depend less and less on the boundary conditions (of course, boundary conditions
are irrelevant in the thermodynamic limit). For $6\times6$--clusters
the difference is down to about 5\%.

\begin{figure}
\epsfxsize=2.8in
\epsffile{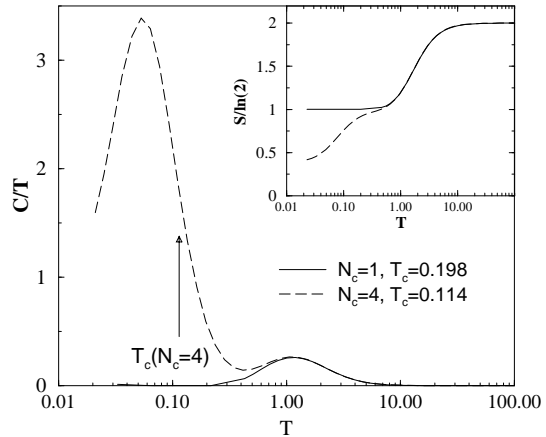}
\caption{Specific heat versus temperature for 1 and 4-site clusters calculated 
with exact enumeration when $U=8$.  For $N_c=1$, there is a single peak
with integrated weight $\ln(2)$ associated with the suppression of
local charge fluctuations. For $N_c=4$, there is an additional peak at
lower temperatures associated with critical fluctuations near the charge 
ordering transition temperature. $T_c$ for $N_c=4$ is indicated
by an arrow.  The entropy $S(T')= \int_0^{T'} dT \frac{C(T)}{T}$
is shown in the inset divided by $\ln(2)$.} 
\label{CoverT}
\end{figure}
\subsection{Energy, entropy and specific heat}
	The DCA differs from the DMFA through the introduction of non-local
dynamical correlations.  For example, in the FKM, the DCA exhibits fluctuations
associated with charge ordering which are absent in the DMFA.  To illustrate 
this, we calculated specific heat divided by the temperature shown in 
Fig.~\ref{CoverT}, using a recently developed maximum-entropy 
method\cite{carey}.  The DMFA ($N_c=1$) result displays a single peak in
$C/T$ associated with the suppression of local charge fluctuations and
the formation of the Mott gap in the single-particle density of states
(Fig.~\ref{localdos}).  As shown in the inset to Fig.~\ref{CoverT}, the 
integrated weight in the peak is $0.69 \approx \ln(2)$; however, the 
infinite temperature entropy $\int_0^\infty \frac{C}{T} dT = 2\ln(2)$ 
for the half filled model.  Thus, only half of the entropy is quenched, 
with the remainder associated with the disorder in $n_f$; i.e.\ $n_f=0$ or 
$n_f=1$ with equal probability on each site when $N_c=1$, regardless of the 
configurations of neighboring sites.  However, when $N_c=4$, $C/T$ displays 
an additional lower-temperature peak slightly below $T=T_c$.  
We believe this 
peak is due to critical fluctuations associated with charge ordering.

\begin{figure}
\epsfxsize=2.8in
\epsffile{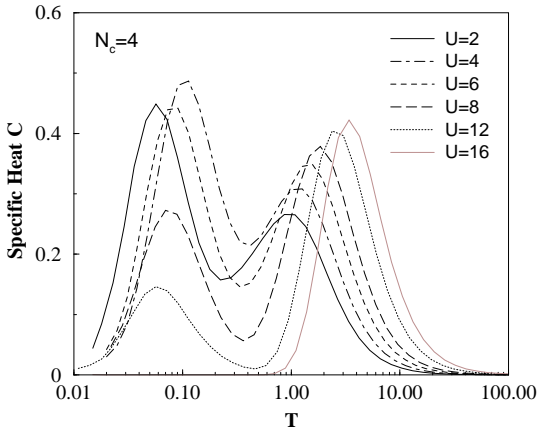}
\caption{Specific heat versus temperature for 4-site custers calculated with 
exact enumeration. The position and height of the lower peak, associated with
charge ordering, is non--monotonic in $U$. For small $U$ the peak rises and 
moves to higher temperatures, for large $U$ the trend is opposite. This 
tracks the behavior of $T_c$ with $U$.  The upper peak, associated with
local (Mott) charge fluctuations, moves higher temperatures and becomes
more pronounced as $U$ increases.} 
\label{C0}
\end{figure}

	To test the identification of the two peaks seen in the DCA
specific heat, we plot $C(T)$ for a variety of values of $U$ when
$N_c=4$ in Fig.~\ref{C0}.  The location of the upper peak increases
monotonically with $U$, consistent with the association of this peak
with local charge flucuations.
However,
the location of the lower peak does not depend monotonically on $U$,
but rather changes in rough proportion to the CDW ordering temperature
shown in Fig.~\ref{phase}. Similar results have been obtained in
Ref. \cite{deraedt}, though we want to point out that in our case the
position of the lower peak is below $T_c$ for the given parameters.
The rise of this lower peak with $U$ for low $U$ (below the maximum $T_c$ and
the opening of the Mott gap) is similar to the half-filled 
Hubbard model \cite{duffy}. 

The total entropy in these lower peaks can be substantial.  For example,
when $U=8$, the entropy $S(T')= \int_0^{T'} dT \frac{C(T)}{T}$ in the
lower peak is $0.41$ whereas that in the upper peak is $0.69\approx\ln{2}$.
Thus, the fluctuations associated with charge ordering quench most 
of the entropy needed to form a proper ground state with $S=0$.

\section{Conclusions} 
We described in detail the recently introduced \cite{fkm} dynamical cluster 
approximation (DCA) and explained its assumptions and approximations.
The DCA systematically introduces non-local corrections to the DMFA.
The DMFA is recovered by taking the cluster to be a single site, 
whereas the exact result is obtained when the cluster becomes large.
We have shown explicitly that the DCA is  causal, systematic, and 
$\Phi$-derivable.  Furthermore, as the cluster size increases, it 
systematically restores momentum conservation neglected in the DMFA.
Consequently, the DCA becomes conserving in the thermodynamic limit.  
We have applied it to an Exact Enumeration and Quantum Monte Carlo study 
of the two-dimensional Falicov-Kimball model and discussed the density of 
states and the spectral function, including their causality and cluster 
size dependence. A pseudo--gap opens in the density of states at 
intermediate interactions as the temperature is lowered, a single--particle 
precursor of the CDW transition at lower temperature.  The phase diagram 
converges monotonically with cluster size, with the notable exception of 
the $2\times2$--cluster. The CDW transition temperature scales linearly in 
the inverse linear dimension of the cluster, as expected for a system in 
the 2D Ising model universality class.  The specific heat clearly displays 
the critical fluctuations associated with  the phase transition, in 
contrast to the dynamical mean field theory where such non--local 
fluctuations are absent.\\

Acknowledgements: It is a pleasure to acknowledge discussions with 
N.E.\ Bickers,
P.G.J.\ van Dongen, 
J. Freericks,
D.\ Hess, 
J.\ Gubernatis, 
M.\ Ma, 
Th.\ Maier,
Th.\ Pruschke,
A. Schiller,
V.\ Sudhindra,
and F.-C. Zhang.
This work was supported by NSF grants DMR-9704021 and DMR-9357199, 
the U.S. Department of Energy, Contract No. W-31-109-ENG-38,
the Ohio Board of Regents Reasearch Challenge Award(HRK),
and the University Grants Commission, India(HRK). 
Computer support provided by the Ohio Supercomputer Center.

\appendix
\section{Two--Particle Properties}
 
Here we discuss the calculation of the lattice two-particle properties, 
such as spin and charge susceptibilities, in terms of the two--particle 
quantities on the cluster. This is a subtle issue which requires some 
formal discussion of what quantities from the cluster and lattice should 
and should not be employed.  We will show using the ``Baym--Kadanoff'' 
formalism that there is a unique construction for which the susceptibities
correspond to the second derivatives of the corresponding extremal free 
energy with respect to external fields.   This optimal choice corresponds
to employing only the irreducible quanties from the cluster when
constructing these susceptibilites.

\subsection{Lattice Quantities and Matrix Notation}

   As discussed in standard texts on quantum many body theory, the charge and
spin susceptibilities at wave vectors $\q$ and frequency $i\nu$ can be 
calculated from the two--particle Green functions $\chi$ as
\begin{eqnarray}
&\left(
\begin{array}{c}
\tchi_{ch}(\q, i\nu) \\
\tchi_{sp}(\q, i\nu) 
\end{array}\right) = & \frac{(k_BT)^2}{N^2} \\
& &\sum_{\k\k' nn', \sigma \sigma'}
\chi_{\q,i\nu,\sigma\sigma'}(\k\;i\omega_n ; \k' i\omega_{n'} )
\left(
\begin{array}{c}
1\\
\sigma\sigma'
\end{array}\right), \nonumber
\end{eqnarray}
where  $\chi$ is the  appropriate Matsubara frequency Fourier \\
$\mbox{component of }
\left< T_{\tau}c^{\dagger}_{\k+\q \sigma}(\tau)c_{\k \sigma}(\tau')
c^{\dagger}_{\k'-\q \sigma'}(\tau'')c_{\k' \sigma'}(\tau''')\right>.$
In diagrammatic perturbation theory, $\chi$ gets related to the 1--particle
irreducible vertex function $\T^{(2)}$ or the particle--hole irreducible 
vertex function $\bgam$ in the standard way as 
\begin{eqnarray}
\bchi_{\q, i\nu}&=& \bchi_{\q,i\nu}^0 + \bchi_{\q,i\nu}^0 \T^{(2)}_{\q,i\nu}
\bchi_{\q,i\nu}^0 \label{diagperta} \\
   &=& \bchi_{\q,i\nu}^0 + \bchi_{\q,i\nu}^0\bgam_{\q,i\nu}\bchi_{\q,i\nu}.
   \label{diagpertb}
\end{eqnarray}
Here, a matrix notation, regarding $\bchi_{\q,i\nu}, \T^{(2)}_{\q,i\nu}$
and $\bgam_{\q,i\nu}$ as matrices with row and column indices labeled by
$(\k\;i\omega_n\sigma)$ and $(\k'\;i\omega_n'\sigma')$ respectively,
has been used to compactify the equations. $(\q\;i\nu)$ constitute passive, 
parametric labels for these matrices.  $\bchi_{\q,i\nu}^0$ is the diagonal
matrix given by
\begin{eqnarray}
\chi_{\q,i\nu,\sigma\sigma'}^0(\k\;i\omega_n; \k' i\omega_{n'})&=&
N\delta_{\sigma\sigma'}\delta_{nn'}\delta_{\k\k'} G_\sigma(\k,i\omega_n)\nonumber \\ 
& & G_\sigma(\k+\q,i\omega_n+i\nu).
\end{eqnarray}
From the above it follows that 
\begin{eqnarray}
\left[\bchi_{\q,i\nu}\right]^{-1}&=&\left[\bchi^0_{\q,i\nu}\right]^{-1} -
\bgam_{\q,i\nu},     \label{susverta}\\
\left[\T_{\q,i\nu}^{(2)}\right]^{-1}&=&\left[\bgam_{\q,i\nu}\right]^{-1}
-\bchi^0_{\q,i\nu}. \label{susvertb}
\end{eqnarray}
For completeness, these
equations may be diagonalized in the spin label to yield the more 
familiar forms
\begin{eqnarray}
\left[\bchi_{\alpha,\q,i\nu}\right]^{-1}&=&\left[\bchi^0_{\q,i\nu}\right]^{-1} -
\bgam_{\alpha,\q,i\nu},     \\
\left[\T_{\alpha,\q,i\nu}^{(2)}\right]^{-1}&=&\left[\bgam_{\alpha,\q,i\nu}\right]^{-1}
-\bchi^0_{\q,i\nu}, 
\end{eqnarray}
where $\alpha$ denotes either the spin or charge channel ($sp$ or $ch$), and
$\Gamma_{sp}=\Gamma_{\sigma,-\sigma}-\Gamma_{\sigma,\sigma}$ and
$\Gamma_{ch}=\Gamma_{\sigma,-\sigma}+\Gamma_{\sigma,\sigma}$.

\subsection{Cluster Quantities}

   On the cluster, the
two--particle Green functions and vertex functions are 
calculated at
the cluster momenta $\Q, \K, \K'$; which we denote by
$\bchi^c_{\Q,i\nu}, \bchi^{0c}_{\Q,i\nu}, \T_{\Q,i\nu}^{(2)c}$ and
$\bgam_{\Q,i\nu}^{c}$, where now the matrix labels correspond to 
$(\K, i\omega_n, \sigma)$ and $(\K', i\omega_n', \sigma')$
(momenta confined to the cluster momenta). These are then
related to each other by the same equations as Eqs.~\ref{susverta} and 
\ref{susvertb}, except that the lattice momenta $\q$ are replaced
by the cluster momenta $\Q$. In a diagrammatic perturbation theory treatment 
of the cluster problem, $\bgam_{\Q,i\nu}^{c}$ is calculated approximately 
as a function of the cluster propagators. In other 
treatments of the cluster, such as QMC, one calculates 
$\bchi^{0c}_{\Q,i\nu}$ and $\bchi^c_{\Q,i\nu}$
and infers $\bgam_{\Q,i\nu}^{c}$ by using the analog of Eq.~\ref{susverta}
in reverse as 
\begin{equation}
\bgam^{c}_{\Q,i\nu}=\left[\bchi^{0c}_{\Q,i\nu}\right]^{-1} -
\left[\bchi^c_{\Q,i\nu}\right]^{-1} \label{inferpi},
\end{equation}
and then  $\T_{\Q,i\nu}^{(2)c}$ using the analog of Eq.~\ref{susvertb}.
Both lattice and cluster quantities are now uniquely defined.

\subsection{Coarse--Grained Quantities}

We now define coarse--grained two--particle Green 
function $ \bar{\chi}$, the equivalent of $\bar{G}$ for the single particle
Green function.
For this purpose, we write $\q=\Q+\tq, \k=\K+\tk,
\k'=\K'+\tk'$, etc, where $\Q, \K, \K'$ are cluster momenta and 
$\tq, \tk, \tk'$ are inside the corresponding momentum cells. 
$ \bar{\chi}$ is then given by
\begin{eqnarray}
 \bar{\bchi}_{\Q+\tq, i\nu} & \equiv & 
\bar{\chi}_{\Q+\tq, i\nu,\sigma\sigma'}(\K, i\omega_n,; \K', i\omega_n')  \\
& = & \frac{N_c^2}{N^2}\sum_{\tk\tk'} 
\chi_{\Q+\tq,i\nu,\sigma\sigma'}(\K+\tk, i\omega_n; \K'+\tk', i\omega_n'), \nonumber 
\end{eqnarray} 
where the first equation again shows the matrix notation.
Similarly $\bar{\bchi}^0_{\Q+\tq, i\nu}$ is the diagonal matrix
with entries given by
\begin{eqnarray}
&&\bar{\chi}^0_{\Q+\tq, i\nu,\sigma\sigma'}(\K, i\omega_n; 
\K', i\omega_n') = N_c\delta_{\sigma\sigma'}\delta_{\K\K'}\delta_{nn'}
\times  \\
&&\left[\frac{N_c}{N}\sum_{\tk}G_\sigma(\K+\tk, i\omega_n)
G_\sigma(\K+\tk+\Q+\tq, i\omega_n+i\nu)\right]. \nonumber 
\end{eqnarray}
For the purposes of calculating 
$\tchi_{ch}(\Q+\tq, i\nu)$ and $\tchi_{sp}(\Q+\tq, i\nu)$, it is enough to 
compute $\bar{\bchi}_{\Q+\tq, i\nu}$, since 
\begin{eqnarray}
&&\left(
\begin{array}{c}
\tchi_{ch}(\Q+\tq, i\nu) \label{cluarr} \\
\tchi_{sp}(\Q+\tq, i\nu) 
\end{array}\right) = \\
&&\frac{(k_BT)^2}{N_c^2}\sum_{\K\K' nn', \sigma \sigma'}
\bar{\chi}_{\Q+\tq,i\nu,\sigma\sigma'}(\K\;i\omega_n; \K' i\omega_{n'}) 
\left(
\begin{array}{c}
1\\
\sigma\sigma'
\end{array}\right). \nonumber 
\end{eqnarray}
For  the single particle Green function we had $\bar{\G}= \G^c$, 
since in that case  the coarse--graining 
is done with the {\it external} momentum.
For the two--particle case, the above defined
 coarse-grained quantities are {\it not} identical
with  $\bchi^c_{\Q, i\nu}$ and $\bchi^{0c}_{\Q, i\nu}$. The  
coarse-grained quantities are defined for all external lattice momenta $\q$, 
not just the cluster momenta $\Q$. 
However, the matrix size is determined
by the number of cluster momenta rather than the (infinite) number 
of lattice momenta.
As we will see below, this is a significant numerical 
simplification, since the
calculation of the susceptibilities can  be reduced to the solution of
a set of linear equations defined on the cluster momenta instead 
of the momenta of the infinite lattice.


\subsection{Two Prescriptions}
Two different prescriptions for computing 
$\bar{\bchi}$ out of cluster quantities 
suggest themselves (a third possibility, approximating
$\bar{\bchi}_{\Q+\tq, i\nu}$ by $\bchi^c_{\Q, i\nu}$, is obviously
too crude to be discussed further). The first one corresponds to replacing
$\mT^{(2)}_{\Q+\tq, i\nu,\sigma\sigma'}
(\K+\tk, i\omega_n; \K'+\tk', i\omega_n')$ by 
$\mT^{(2)c}_{\Q, i\nu,\sigma\sigma'}(\K, i\omega_n; \K', i\omega_n')$
in the expression for $\bar{\bchi}_{\Q+\tq, i\nu}$ derived
from Eq.~\ref{diagperta}. We then get the equation
\begin{equation}
\bar{\bchi}_{\Q+\tq, i\nu} \cong \bar{\bchi}^0_{\Q+\tq, i\nu} +
\bar{\bchi}^0_{\Q+\tq, i\nu}\T^{(2)c}_{\Q, i\nu}\bar{\bchi}^0_{\Q+\tq, i\nu}.
\label{firpres}
\end{equation} 
This means we have identified the {\it reducible} 
two--particle vertex $\T^{(2)}$
of the cluster and the lattice at the cluster momenta.

The second prescription, that we argue below is the {\it correct} 
prescription, is to replace 
$\Gamma^{(2)}_{\Q+\tq, i\nu,\sigma\sigma'}(\K+\tk, i\omega_n; \K'+\tk', i\omega_n')$ by
$\Gamma^{(2)c}_{\Q, i\nu,\sigma\sigma'}(\K, i\omega_n; \K', i\omega_n')$    
in the integral equation for $\bar{\bchi}_{\Q+\tq, i\nu}$ derived
from Eq.~\ref{diagpertb}. This leads to the equation
\begin{equation}
\bar{\bchi}_{\Q+\tq, i\nu} \cong \bar{\bchi}^0_{\Q+\tq, i\nu} +
\bar{\bchi}^0_{\Q+\tq, i\nu}\bgam^{c}_{\Q, i\nu}\bar{\bchi}_{\Q+\tq, i\nu},
\end{equation} 
whence 
\begin{equation}
\bar{\bchi}_{\Q+\tq, i\nu}=\left(\left[\bar{\bchi}^0_{\Q+\tq, i\nu}\right]^{-1}
-\bgam^{c}_{\Q, i\nu}\right)^{-1}.
\label{secpres}
\end{equation}
Here, we have identified the {\it irreducible} two--particle vertex $\bgam$
of the cluster and the lattice at the cluster momenta.
Either Eq. \ref{firpres} or  Eq.  \ref{secpres} 
can then be used in Eq. \ref{cluarr} to compute $\tchi_{ch}$ and 
$\tchi_{sp}$. 
At this stage  it is not clear which prescription is better or whether
both could be feasible approximations. We will now show that
internal consistency and $\Phi$--derivability in the Baym--Kadanoff 
sense do single out the second prescription, Eq. \ref{secpres}.

\subsection{Relation to $\Phi$-derivability}

The Baym--Kadanoff\cite{baym} $\Phi$  functional is diagrammatically
defined as
\begin{equation}
\Phi({\bf{G}})=\sum_l p_l \mbox{tr}\left[{\bf{\Sigma}}_\sigma^l
{\bf{G}}_\sigma\right].
\label{phidef}
\end{equation}
The trace indicates summation over frequency, momentum and spin. Here, 
${{\bf{\Sigma}}_\sigma}^l$ is the set of irreducible self energy diagrams
of $l^{th}$ order in the interaction, ${\bf{G}}_\sigma$ is the dressed Green 
function related to ${{\bf{\Sigma}}_\sigma}$ and the bare lattice Green 
function ${\bf{G}}_\sigma^0$ via the Dyson equation 
${\bf{G}}_\sigma^{-1} = {\bf{G}}_\sigma^{0-1} - {{\bf{\Sigma}}_\sigma}$, 
and $p_l$ is a counting factor equal to the number of occurrences of 
${\bf{G}}_\sigma$ in each term (for Hubbard-like models, $p_l=1/l$).
The free energy  $\Im$ can be expressed in terms of the ``linked cluster
expansion'' $W$ as  $\Im= -k_B T\ W$ with 
\begin{equation}
W=\Phi({\bf{G}})-\mbox{tr}\left[{\bf{\Sigma}}_\sigma {\bf{G}}_\sigma\right] 
-\mbox{tr}\ln\left[-{\bf{G}}_\sigma\right].
\label{wdef}
\end{equation}
With the above  definitions it holds that 
${\bf{\Sigma}}_\sigma=\delta\Phi/\delta {\bf{G}}_\sigma$, as required for a 
``$\Phi$--derivable'' theory, and the free energy is stationary
under variations of ${\bf{G}}$.
In addition, the irreducible vertex function is obtained by a second
variation of $\Phi$, 
${\bf{\Gamma}}_{\sigma,\sigma'}= 
\delta^2\Phi/(\delta {\bf{G}}_\sigma \delta {\bf{G}}_{\sigma'}) = 
\delta{\bf{\Sigma}}_\sigma/\delta {\bf{G}}_{\sigma'}$.

The DCA can be microscopically motivated by our  choice of the Laue 
function $\Delta_{DCA}$ in Eq. \ref{laue_DCA}. The effect of the chosen
Laue function is the replacement of the ${\bf{\Sigma}}_\sigma$ and 
${\bf{\Gamma}}_{\sigma,\sigma'}$ by the
corresponding coarse--grained quantities (indicated by the bars).
For example, consider the relation $\bSig = \mT^{(2)} {\bf{G}}$ 
(order by order in the diagrammatic series ).
The vertices connecting the Green function to $\mT^{(2)}$ 
do not preserve momentum within the cells about the
cluster momentum due to the DCA Laue function. Consequently, the lattice 
Green function ${\bf{G}}_\sigma$  is replaced by the coarse--grained Green function
$\bar{{\bf{G}}}_\sigma$. The external momentum label ($\k$) of the 
self energy is in principle still a lattice momentum, however, the self energy
will only depend through the function $\M(\k)$ on $\k$.
If  we use this self energy in, e.g., the calculation of its 
contribution to the $\Phi$ functional, the Laue function on the vertices
will ``reduce'' both the self energy as well as the diagram closing Green 
function to their corresponding coarse--grained expressions. Consequently,
the DCA $\Phi$ functional reads
\begin{equation}
\Phi_{DCA}({\bf{G}})=
\sum_l p_l
\mbox{tr}\left[\bar{{\bf{\Sigma}}}_\sigma^l\bar{{\bf{G}}}_\sigma\right].
\label{phidca} \end{equation}
In correspondence to the lattice system, 
\begin{equation}
\frac{\delta\Phi_{DCA}}{\delta\bar{{\bf{G}}}_\sigma}=\bar{{\bf{\Sigma}}}=
\frac{\delta\Phi_{DCA}}{\delta{{\bf{G}}_\sigma}},
\label{identify}
\end{equation}
where the second equality follows since the variation 
$\delta/\delta {\bf{G}}_\sigma$
corresponds to cutting a Green function line, so that 
$\delta\bar{{\bf{G}}}_{\sigma \K}/\delta {{\bf{G}}}_{\sigma' \k'}= 
\delta_{\K,\M(\k')}\delta_{\sigma,\sigma'}$.  
It follows that the DCA estimate of the lattice
free energy 
is $\Im_{DCA}= -k_BT\ W_{DCA}$, where
\begin{equation}
W_{DCA}= \Phi_{DCA}-\mbox{tr}\left[{\bf{\Sigma}}_\sigma {\bf{G}}_\sigma\right] 
-\mbox{tr}\ln\left[-{\bf{G}}_\sigma\right].
\end{equation}
$W_{DCA}$ is stationary with respect to ${\bf{G}}_\sigma$,
\begin{equation}
\delta\Im_{DCA}/\delta {\bf{G}}_\sigma=
-\bar{{\bf{\Sigma}}}_\sigma+{\bf{\Sigma}}_\sigma=0,
\end{equation}
which means that $\bar{{\bf{\Sigma}}}_\sigma$ is the proper approximation for 
the lattice self energy corresponding to $\Phi_{DCA}$. 

The susceptibilities are thermodynamically defined as second derivatives 
of the free energy with respect to external fields.  $\Phi_{DCA}({\bf{G}})$
and $\bar{{\bf{\Sigma}}}_\sigma$, and hence $\Im_{DCA}$ depend on these fields 
only through ${\bf{G}}_\sigma$ and ${\bf{G}}_\sigma^0$.  Following Baym\cite{baym} it is 
easy to verify that, the prescription  (\ref{cluarr}$+$\ref{secpres}), with 
\begin{equation}
{\bf{\Gamma}}_{\sigma,\sigma'}\approx \bar{{\bf{\Gamma}}}_{\sigma,\sigma'} 
\equiv \delta\bar{{\bf{\Sigma}}}_\sigma/\delta {\bf{G}}_{\sigma'}.
\end{equation}
yields the same estimate that would be obtained from the second derivative 
of $W_{DCA}$ with respect to the applied field.
For example, the first derivative of the partition function $W_{DCA}$ with 
respect to a spatially homogeneous external magnetic field $h$ is
the magnetization,
\begin{equation}
m=\mbox{tr}\left[\sigma {\bf{G}}_\sigma\right].
\end{equation}
The susceptibility is given by the second derivative, 
\begin{equation}
\frac{\partial m}{\partial h}=
\mbox{tr}\left[\sigma\frac{\partial {\bf{G}}_\sigma}{\partial h}\right].
\end{equation}
We substitute ${\bf{G}}_\sigma =
\left( {\bf{G}}_\sigma^{0-1}- \bar{{\bf{\Sigma}}}_\sigma\right)^{-1}$, and
evaluate the derivative, 
\begin{equation}
\frac{\partial m}{\partial h}=
\mbox{tr}\left[\sigma\frac{\partial {\bf{G}}_\sigma}{\partial h}\right]
=\mbox{tr}\left[
{\bf{G}}_\sigma^2
\left( 1 +
\sigma\frac{\partial \bar{{\bf{\Sigma}}}_\sigma}{\partial {\bf{G}}_{\sigma'}}
  \frac{\partial {\bf{G}}_{\sigma'}}{\partial h}
\right)
\right],
\end{equation}
where $\frac{\partial m}{\partial h}=\tchi_{sp}(\q=0, i\nu=0)$.  If we identify 
$\chi_{\sigma,\sigma'}=\sigma \frac{\partial{\bf{G}}_\sigma}{\partial h}$,
and $\chi_{\sigma}^0= {\bf{G}}_\sigma^2$, collect all of the terms 
within both traces, and sum over the cell momenta $\tk$, we obtain the 
two--particle Dyson's equation 
\begin{eqnarray} 
2\big({\bar{\chi}}_{\sigma,\sigma}&-&{\bar{\chi}}_{\sigma,-\sigma}\big)\\
&=&
2{\bar{\chi}}_{\sigma}^0 +
2{\bar{\chi}}_{\sigma}^0 \left( \bar{{\bf{\Gamma}}}_{\sigma,\sigma} 
-\bar{{\bf{\Gamma}}}_{\sigma,-\sigma}\right)
\left({\bar{\chi}}_{\sigma,\sigma}-{\bar{\chi}}_{\sigma,-\sigma} \right)\nonumber
\end{eqnarray}
which is equivalent to Eq.~A15.  
We see that indeed it is the irreducible quantity, i.e. the
vertex function, for which cluster and lattice correspond.

In summary, the choice of the Laue function and the 
requirement of a $\Phi$-derivable theory
ultimately determine the way lattice properties are constructed 
out of cluster properties.
The usefulness of the DCA lies in the fact that
both the single and the two--particle irreducible properties ($\bar{\bSig}$ and
$\bar{\bgam}$) can be  determined from  the cluster problem, i.e.
$\bar{\bSig}={\bSig}^c$ and $\bar{\bgam}=\bgam^c$.
Note that, although this construction is unique and $\Phi$-derivable,
because  of  the partial violation of momentum conservation at each internal
vertex  described by $\Delta_{DCA}$ certain Ward identities will be violated
in  any dimension, even for the single site cluster (DMFA) appropriate in 
D=$\infty$.  This will be discussed in Appendix D.

\section{DCA for Problems with extended range or electron-phonon
interactions}
%
In this Appendix we present an extension of the DCA to problems with
extended range interactions, such as in the extended Hubbard Model.

Consider the partition function for such a model written in terms of
Fermionic functional integrals: 
\begin{eqnarray}
\mathcal{Z} &=\int_{cc^{\dagger }}\exp-\int_{0}^{\beta }d\tau&
\Big[\sum_{ij}c_{i}^{\dagger }(\tau )\{(\partial _{\tau }-\mu )\delta
_{ij}-t_{ij}\}c_{j}(\tau) \nonumber \\ 
&&+\ U\ \sum_{i}\hat{n}_{i\uparrow }(\tau
)\hat{n}_{i\downarrow }(\tau )  \\
&&+\ \frac{1}{2}\ \sum_{i\neq j}
\sum_{\sigma \sigma ^{\prime }}V_{ij}\hat{n}_{i\sigma
}(\tau )\hat{n}_{j\sigma ^{\prime }}(\tau )\Big]. \nonumber
\end{eqnarray}

By introducing a real, continuous Hubbard -Stratonovich field 
$\phi_{i}(\tau )$ which couples to the local charge density $\hat{n}_{i}\equiv
\sum_{\sigma }\hat{n}_{i\sigma }$ , we can write 
\begin{eqnarray}
\mathcal{Z} &=&\int_{cc^{\dagger }}\int_{\phi }exp-\int_{0}^{\beta }d\tau
\Big[\sum_{ij}c_{i}^{\dagger }(\tau )\{(\partial _{\tau }-\tilde{\mu})\delta
_{ij}-t_{ij}\}c_{j}(\tau )  \nonumber \\
&&+\widetilde{U}\sum_{i}n_{i\uparrow }(\tau )n_{i\downarrow }(\tau )+\frac{
\tilde{V_{o}^{2}}}{2}\sum_{ij}\phi _{i}(\tau )(\widetilde{V}
)^{-1}{}_{ij}\phi _{j}(\tau )  \nonumber \\
&&+\widetilde{V}_{o}\sum_{i}\phi _{i}(\tau )\hat{n}_{i}(\tau )\Big].
\end{eqnarray}
Here, $\tilde{V}_{ij}=\tilde{V}_{o}\delta _{ij}-V_{ij}$ with $\tilde{V}_{o}$
so chosen as to make $\tilde{V}$ positive definite (and hence invertible), $
\tilde{U}=(U+\tilde{V}_{o})$ and $\tilde{\mu}=\mu -\frac{1}{2}\tilde{V}_{o}$.
 For example, for the extended Hubbard Model with nearest neighbor
interaction of strength $V$, $\tilde{V}_{o}=zV$, where $z$ is the
co-ordination number of the lattice.

Now it is straightforward to devise the DCA for this coupled Fermion-Boson
problem. The cluster problem we need to solve corresponds to the functional
integral given by 
\begin{eqnarray}
Z_{c} &=&\int_{cc\dag }\int_{\phi }\exp -\Big[\int_{0}^{\beta }d\tau
\int_{0}^{\beta }d\tau ^{\prime }\sum_{ij}\{c_{i}^{\dag }(\tau ){\mathcal{G}}
_{ij}^{-1}(\tau -\tau ^{\prime })c_{j}(\tau ^{\prime })  \nonumber \\
&&+\phi _{i}(\tau ){\mathcal{D}}_{ij}^{-1}(\tau -\tau ^{\prime })\phi
_{j}(\tau ^{\prime })\}  \\
&&+\int_{0}^{\beta }d\tau \sum_{i}\{\tilde{U}\hat{n}_{i\uparrow }(\tau )
\hat{n}_{i\downarrow }(\tau )+\tilde{V}_{0}\phi _{i}(\tau )
\hat{n}_{ij}(\tau )\}\Big].\nonumber 
\end{eqnarray}
The cluster problem is to be treated by some technique to obtain the cluster
propagators and self energies: $G^{c}(\K)$ , $\Sigma ^{c}(\K)$ for the
electrons and $D^{c}(\Q)$, $\Pi ^{c}(\Q)$ for the field $\phi $ , at cluster
momenta $\K$ and $\Q$ . One has the Dyson equations:
\begin{eqnarray}
\left( G^{c}(\K)\right) ^{-1} &=&{\mathcal{G}}^{-1}(\K)-\Sigma ^{c}(\K)\ , \\
\left( D^{c}(\Q)\right) ^{-1} &=&{\mathcal{D}}^{-1}(\Q)-\Pi ^{c}(\Q)\ ,
\end{eqnarray}
where the frequency arguments have been suppressed for convenience.

The self--consistent embedding of the above cluster in the effective medium
defined by the rest of the sites of the original lattice is obtained by
assuming that $\Sigma ^{c}(\K)$ , and $\Pi ^{c}(\Q)$ represent good
approximations to the (coarse--grained averages of the ) lattice self
energies, and that $G^{c}(\K)$ and $D^{c}(\Q)$ must equal the coarse--grained
averages of the corresponding lattice green functions: 
\begin{eqnarray}
G^{c}(\K) &=&\bar{G}(\K)\equiv \sum_{\tilde{\k}}\frac{1}{i\omega_n+
\tilde{\mu}-\epsilon_{\tilde{\k}+\K}-\Sigma^c(\K)}\ , 
\\
D^c(\Q) &=&\bar{D}(\Q)\equiv \sum_{\tilde{\q}}
\frac{1}{\tilde{V}_{\Q+\tilde{\q}}^{-1}-\Pi^c(\Q)}.
\end{eqnarray}

Thus, the self consistency loop is closed by recalculating ${\mathcal{G}}
_{\K}^{-1}$ and ${\mathcal{D}}_{\Q}^{-1}$ using the Dyson equations backwards as 
\begin{eqnarray}
{\mathcal{G}}^{-1}(\K) &=&\bar{G}^{-1}(\K)+\Sigma ^{c}(\K)\ , \\
{\mathcal{D}}^{-1}(\Q) &=&\bar{D}^{-1}(\Q)+\Pi ^{c}(\Q).
\end{eqnarray}

We note that for the 1-site cluster, the resulting DMFA does not correspond
to the approximation resulting from scaling $V$ as $\frac{V^{*}}{d}$ (whence
in the $D\rightarrow \infty $ limit only the Hartree contribution to $\Sigma $
survives), but is a rather different approximation which includes local
dynamical charge fluctuations and local screening effects\cite{smith-si}. 
It is formally similar to the problem
obtained in the DMFA of the Holstein - Hubbard model. Correspondingly, the
DCA for this latter model can be formulated analogously to the
above.

\section{Proof of causality}

In this Appendix we prove that the DCA formally preserves the
condition of positive semi-definiteness of the single-particle
spectral functions. The proof requires
that the cluster problem is solved by methods that preserve causality
(exact enumeration, QMC, etc.).  For simplicity of notation the proof 
is explicitly given for Hubbard like models, but it can be easily
generalized to the PAM, multi--band models and models with
non--local interactions.

Most steps of the DCA algorithm are easily seen to preserve the causality
property.  We assume a causal ${\cal{G}}$, so that $-{\rm{Im}} {\cal{G}}>0$,
as a starting point of the
iteration. If the method to solve the cluster problem preserves causality
the resulting cluster Green function $G^c$ will also be causal. With
Dyson's equation we obtain a causal cluster self energy. This self energy
is also assumed to be the lattice self energy of the infinite lattice 
at the cluster momenta. Therefore, the  lattice Green function ( the summand
of Eq. \ref{gbar}) is also causal. As the coarse--grained Green function
 $\bar{G}$ is obtained by an average of causal Green functions 
it must be causal, too.

The nontrivial step is to show that Eq. \ref{gsck} does not lead to
an acausal ${\cal{G}}$ for the next iteration. The spectral function
of  ${\cal{G}}$ will be positive semidefinite if 
\begin{equation}
\mbox{Im} (\bar{G}(\K,\omega)^{-1}) \geq -{\rm Im}\Sigma^c(\K,\omega)\,.
\label{cond}
\end{equation}
We write $\bar{G}(\K,\omega)$ as
$\bar{G}(\K,\omega)=(N_c/N) \sum_{\tk} (z_{\K+\tk}(\omega))^{-1}$ 
with $z_{\K+\tk}(\omega)=x_{\tk}(\K,\omega)+ia(\K,\omega)$.  
$z_{\K+\tk}(\omega)$ is the inverse of our
estimate of the Green function of the infinite lattice with a real part
$x_{\tk}(\K,\omega)=\omega -\epsilon_{\K +\tk}-{\rm{Re}}\Sigma^c(\K,\omega)$ 
and an imaginary part $a(\K,\omega)=-{\rm Im}\Sigma^c(\K,\omega)$, 
with $a(\K,\omega)$ a  positive semidefinite function of $\K$ and $\omega$ 
but independent of $\tk$. Graphically, the proof of Eq. \ref{cond} is
illustrated  in Fig. \ref{causal}.

We now proceed to show the validity of Eq. \ref{cond} in a rigorous fashion.
To simplify notation we will suppress the common indices $\K$ and $\omega$. 
We also specify to the retarded Green functions with 
$\omega \rightarrow \omega +\imath \eta$ with positive infinitesimal $\eta$. 
The sum over $\tk$ in the definition of $\bar{G}$
runs over $n=N/N_c$ terms. Each term is 
a complex number with a positive definite  
imaginary part $a$ that is {\it independent} of the summation index.
Eq. \ref{cond} is now cast into the following proposition:\\

{\bf Proposition:}
For $j=1, \ldots, n$, let $z_{j} \in {\bf C}$, where ${\bf C}$ is the set 
of complex numbers, and
 $\mbox{Im}\ (z_{j}) = a > 0$.  If 
$$\bar{G} :=\frac{1}{n} \sum_{j=1}^{n}
\frac{1}{z_j}, \,\,\,\mbox{then} \,\,\, 
 \mbox{Im}\,\left({\bar{G}}^{-1}\right) \geq a\, , $$ with
equality if and only if
$z_{1} = \cdots = z_{j}=\cdots = z_{n}  ~$. 

{\it Proof:}
If $w = u + i v = \frac{1}{z}$, with $z =
x + iy$, then the line $\mbox{Im}\ z = a$,
in the extended $z$-plane,  given by
$$\mbox{Im} (z) = y = a = \frac{-v}{u^{2} + v^{2}},$$
is mapped, in a one-to-one fashion, onto the circle
$$u^{2} + \left(v + \frac{1}{2a}\right)^{2} =
\left(\frac{1}{2a}\right)^{2}$$
in the extended $w$-plane, with center
$-i/2a$ and
a radius of  $r= 1/2a$.
It follows that $\displaystyle \bar{G}$ lies on or
inside this circle:
\begin{eqnarray}
\left|\bar{G} - \left(\frac{-i}{2a}\right)\right|& =&
\frac{1}{n}\left| \sum_{j=1}^{n} \left(\frac{1}{z_{j}}
+ \frac{i}{2a}\right)\right|  \nonumber \\
& \leq & \frac{1}{n} \sum_{j=1}^{n}
\left| \frac{1}{z_{j}} + \frac{i}{2a}\right| = \frac{1}{2a}\, ,
\end{eqnarray}
where we have used the triangular inequality.
The bijective function
$z=1/w$  maps a point $w$ 
strictly inside the circle to a point $z$ with
{\rm Im}$(z) > a$ (and conversely):
$$\mbox{Im}\ z = \frac{-v}{u^{2}+v^{2}} > a$$
if and only if
$$u^{2} +\left(v+\frac{1}{2a}\right)^{2} < \left(\frac{1}{2a}\right)^{2}.$$
Hence, {\rm Im}$\left({\bar{G}}^{-1}\right) \geq a$, where equality
holds if and only if $z_{1} = \cdots = z_{j}=\cdots = z_{n}  ~$.\\

Because of the infinitesimal $\eta$ we had $a > 0$ for the above proof.
However, if ${\rm Im} \Sigma^c(\K,\omega) = 0$, the resulting imaginary part
of ${\cal{G}}$ is proportional to $-\eta$.  This is the case, e.g.,
for frequencies larger than the band width. Hence, the band width of
${\cal{G}}$ is identical to the band width of $\bar{G}$ and $G^c$, i.e.
there is no band broadening induced by the coarse--graining procedure.

Generalization to multiband models such as the PAM is straightforward. 
Without going into the details of the  model we note that there are 
two species of fermions which are coupled by  on--site hybridization. 
The d--electrons are itinerant and noninteracting, whereas the f--electrons 
are localized (no bare hopping) and have 
a Hubbard interaction. The f--electron Green function has two self energies,
from the Hubbard interaction and the hybridization, respectively. Both
self energies are causal (negative semidefinite and decaying like $1/\omega$).
In contrast to the Hubbard self energy
the self energy due to the hybridization is known explicitly and does
depend on all the lattice momenta, therefore also on the $\tk$ 
momenta in the cells about the cluster momenta. 
For a given $\K$ and $\omega$ the imaginary part of this
self energy is bounded from above 
by some  value $-b_{\rm min}(\K,\omega)$. Consequently, we can prove
in analogous fashion that 
$${\rm Im} (\bar{G}_f(\K,\omega)^{-1}) \geq a(\K,\omega)+b_{\rm min}(\K,\omega)\, ,$$
where $-a(\K,\omega)$ is the self energy due to the Hubbard interaction of the
f--electrons.

A last remark on the possibility of self energy interpolation is in order 
here.  At first glance one might try to improve the calculation by
employing an interpolation of the cluster self energy between the cluster 
momenta in the coarse graining step Eq. \ref{gbar}, rather than using the 
``rectangular'' approximation for the lattice self energy 
$\Sigma(\K+\tk,\omega)\approx \Sigma^c(\K,\omega)$. However, as 
one can easily convince oneself given  the above proof, {\it any} interpolation 
scheme will violate causality if ${\rm Im}\Sigma^c(\K,\omega)$ has a minimum 
somewhere in the BZ. This will generally
be so  except in the case of the single site cluster, in which there
is nothing to interpolate. This further limits the
freedom  of the coarse--graining procedure.

\section{Conservation of the DMFA and DCA}

	An approximation which satisfies the various Ward identities is 
identified as a ``conserving approximation'' since the Ward identities are
derived from conservation laws.  Baym and Kadanoff\cite{baymandkad,baym} 
showed that a sufficient condition to guarantee that an approximation is 
conserving is for it to be $\Phi$-derivable and self--consistent.  Energy, 
particle number, and momentum are also assumed to be conserved at each 
internal vertex, which may be assured by properly constructing the diagrams 
from the lattice propagator $G_\k$ using well-known Feynman rules.
Specifically, the functional $\Phi\left(G(\k,\omega),U\right)$ is a set of 
closed graphs formed from the lattice propagators $G(\k,\omega)$ and 
interactions $U$.    The one- and two--particle self energies are calculated 
from functional derivatives of $\Phi(G(\k,\omega),U)$, 
$\Sigma(\k,\omega)=\delta\Phi/\delta G(\k,\omega)$, 
$\bgam_{\sigma,\sigma'}=\delta^2\Phi/\delta \G_\sigma\delta \G_{\sigma'}$.  
The equation $\Sigma(\k,\omega)=\delta\Phi/\delta G(\k,\omega)$ must be 
solved self--consistently until $G(\k,\omega)$ converges.  As an additional
consequence, Baym showed that quantities calculated within such an
approximation were unique.

In the infinite-dimensional formalism of Metzner and Vollhardt momentum
conservation is violated at internal vertices.  Consequently, $\Phi$ is 
a functional of the local propagator $G_{ii}(\omega)$ rather than the lattice 
propagator $G({\k},\omega)$, and the corresponding self energies are obtained
from functional derivatives of $\Phi(G_{ii}(\omega),U)$ and are therefore
also local.  However, we may also expect violations of some conservation
laws.  If a proper $\Phi(G(\k,\omega),U)$ is taken, all non-local diagrams 
which are higher order in $1/D$ vanish, so that 
$\Phi(G(\k,\omega),U)=\Phi(G_{ii}(\omega),U) +{\cal{O}}(1/D)$.  
Each functional 
derivative with respect to the Green function breaks an internal line 
and so reduces the order of the approximation by 
$\sqrt{D}$\cite{muller-hartmann}.  
It follows then that the self energy is also local
$\delta\Phi(G(\k,\omega),U)/\delta G(\k,\omega) = 
\Sigma(G(\k,\omega),U)=\Sigma(G_{ii}(\omega),U)+{\cal{O}}(1/\sqrt{D})$.
However a problem emerges at the two--particle, or higher, level
since $\Gamma(G(\k,\omega),U)=\Gamma(G_{ii}(\omega),U)+{\cal{O}}(1)$ for any $D$,
with the difference due to needed non-local corrections.
Equivalently, if $\Phi$ is evaluated in the limit $D\to\infty$ before 
the functional derivatives are evaluated, then 
$\Gamma(G(\k,\omega),U)=\Gamma(G_{ii}(\omega),U)$; however, if the order is reversed,
then corrections of order unity are required\cite{PVD_note}.  Thus, 
due to the lack of momentum conservation, the DMFA does not provide
a unique prescription for the calculation of two--particle properties
and thus it need not be conserving.

\begin{figure}
\epsfxsize=2.6in
\epsffile{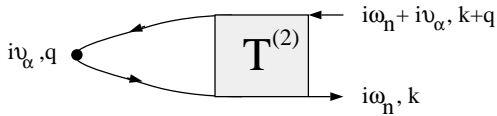}
\caption{Definition of $i\nu_\alpha\Lambda_0$ and ${\q\cdot\bf{\Lambda}}$.
Here, each solid line is a full lattice propagator $G(\k,\omega)$,
the filled box is the full particle-hole reducible two--particle
T-matrix, and the filled circle $\bullet$ is $i\nu_\alpha$ or
$\epsilon_{\k+\q} -\epsilon_\k$ for $i\nu_\alpha\Lambda_0$ or
${\q\cdot\bf{\Lambda}}$, respectively.} 
\label{ward_fig1}
\end{figure}
For example, the equation of continuity, 
$\grad\cdot \J -\partial \rho/\partial t =0$, which describes charge 
conservation by electric currents, yields the original Ward\cite{ward} 
identity
\begin{equation}
i\nu_\alpha \Lambda_0-\q\cdot{\bf{\Lambda}} 
= 
\Sigma( \k+\q,i\nu_\alpha +i\omega_n) -\Sigma(\k,i\omega_n),
\label{ward_cont}
\end{equation}
where $\Lambda_0$ and ${\bf{\Lambda}}$ are the scalar and vector
components of the dressed vertex function such that 
\begin{eqnarray}
\Lambda_0(\k,\q,i\omega_n,i\nu_\alpha) &=& 
\frac{T}{N} \sum_{\k',n'} G(\k',i\omega_n')
G(\k'+\q,i\omega_n'+i\nu_\alpha)\nonumber \\
&&\mT^{(2)}_{\q, i\nu_\alpha}(\k,i\omega_n;\k',i\omega_n')
\label{lambda0}
\end{eqnarray}
and 
\begin{eqnarray}
\q\cdot{\bf{\Lambda}}(\k,\q,i\omega_n,i\nu_\alpha) &=&
\frac{T}{N} \sum_{\k',n'} \left(\epsilon_{\k'+\q}-\epsilon_{\k'}\right)
G(\k',i\omega_n')\nonumber\\
&&G(\k'+\q,i\omega_n'+i\nu_\alpha)\nonumber\\
&&\mT^{(2)}_{\q, i\nu_\alpha}(\k,i\omega_n;\k',i\omega_n') .
\label{lambdavec}
\end{eqnarray}
Here, $\T^{(2)}$ is the corresponding particle-hole reducible two--particle 
T-matrix $$\T^{(2)}_{\q,i\nu_\alpha} = \bgam^{ph}_{\q,i\nu_\alpha}
\left(1-\bchi_{\q,i\nu_\alpha}^0\bgam^{ph}\right)^{-1},$$ and
$\bgam_{ph}=\bgam_{\sigma,\sigma}+\bgam_{\sigma,-\sigma}$ is the 
particle-hole irreducible two--particle self energy, 
with $(\k, i\omega_n)$ and $(\k', i\omega_n')$ as the matrix indices,
and $\bchi^0$ is the diagonal matrix with entries
$\chi_{\q,i\nu_\alpha}^0(i\omega_n,i\omega_n')\equiv N\delta_{nn'}
\delta_{\k\k'} G(\k,i\omega_n) G(\k+\q,i\omega_n+i\nu_\alpha)$, and 
$\epsilon_{\k'}$ the bare electronic dispersion.
The corresponding diagrams are illustrated in Fig.~\ref{ward_fig1}.

When this formalism is applied as the DMFA in finite dimensions, the 
conservation of Ward identities does not follow from the arguments of 
Baym and Kadanoff.  If we write down a proper $\Phi(G(\k,\omega),U)$, 
the only way to obtain the local generating 
function $\Phi(G_{ii}(\omega),U)$ used in the DMFA is to ignore momentum 
conservation within each graph and sum over each internal momenta 
independently.  This clearly violates the requirement for a conserving 
approximation that momentum be conserved at each internal vertex\cite{baym}, 
so the conserving property of the theory is lost.  

This can be seen from a direct examination of Ward's original identity;  
i.e., the Ward identity Eq.~\ref{ward_cont} is not satisfied for a general 
${\q}$ except when $i\nu _{\alpha }$ is zero. To see this, note that 
from Eq.~\ref{lambda0} and Eq.~\ref{lambdavec} and some simple algebra one 
can write
\begin{eqnarray}
i\nu _{\alpha }\Lambda _{0}&-&{\q}\cdot {\bf {\Lambda }} = \nonumber\\
&&\frac{T}{N}
\sum_{{\k}^{\prime },n^{\prime }}\LARGE[\{G(\k^\prime,i\omega_{n}^{\prime})-
G(\k'+\q,i\omega_n^\prime+i\nu _{\alpha })\}+  \nonumber \\
&&\{\Sigma ( \k^\prime+\q, i\nu_{\alpha}+i\omega _{n})
-\Sigma (\k^\prime,i\omega_n )\} \nonumber \\
&&G(\k^\prime, i\omega_n^\prime )
G(\k^\prime + \q , i\omega_n^\prime +i\nu _{\alpha})\LARGE]  \nonumber \\
&&\mT^{(2)}_{\q, i\nu_\alpha}(\k,i\omega_n;\k',i\omega_n').
\label{ward_rewrite}
\end{eqnarray}
Specializing now to the DMFA, the required Ward identity can be written as
\begin{eqnarray}
\Sigma (i\nu _{\alpha }&+&i\omega _{n})-\Sigma (i\omega _{n}) = \nonumber \\
&&\frac{T}{N}
\sum_{{j},n^{\prime }}\LARGE[\{G_{ii}(i\omega _{n}^{\prime })-G_{ii}(i\omega
_{n}^{\prime }+i\nu _{\alpha })\}\delta _{ij}  \nonumber \\
&&+ \{\Sigma (i\nu _{\alpha }+i\omega _{n})-\Sigma (i\omega
_{n})\}  \exp(i\q\cdot\r_{ij})\nonumber \\
&&G_{ij}(i\omega _{n}^{\prime })G_{ji}(i\omega _{n}^{\prime }+i\nu
_{\alpha })\LARGE]\nonumber \\
&& \mT^{(2)}_{\q, i\nu_\alpha}(i\omega _n,i\omega _n'),
\label{ward_dmfa}
\end{eqnarray}
where we have used the DMFA in the second step and assumed that $\Sigma $
and $\bgam^{ph}$ are momentum independent, so 
$\T^{(2)}_{\q, i\nu_\alpha} = \bgam^{ph}_{i\nu_\alpha}\left(1-
\bchi_{\q,i\nu_\alpha}^0\bgam^{ph}_{i\nu_\alpha}\right)^{-1}$
has only the momentum dependence it inherits from $\bchi_{\q,i\nu_\alpha}^0$. 
Clearly, when $i\nu _{\alpha }$ is
zero, the RHS vanishes for arbitrary ${\bf {q}}$ and the Ward identity is
satisfied. But when $i\nu _{\alpha }$ is nonzero, the second term on the
right hand side has a nontrivial ${\bf {q}}$ dependence in general, and the
Ward identity is violated since the LHS of Eq.~\ref{ward_dmfa} is 
${\bf {q}}$ independent.

Even in the $D\rightarrow\infty$ limit the Ward identity is not always
satisfied.  From the form of Eq.~\ref{ward_dmfa} it is clear that
the Ward identity is only satisfied when 
$\chi_q^0(i\omega_n,i\nu_\alpha)\equiv \frac{1}{N} \sum_k  G(\k,i\omega_n)
G(\k+\q,i\omega_n+i\nu_\alpha) = \chi_{ii}^0 (i\omega_n,i\nu_\alpha)
\equiv G_{ii}(i\omega_n)G_{ii}(i\omega_n+i\nu_\alpha)$.  This is true for 
a generic ${\bf {q}}$ where  
$X(\q)=\frac{1}{D}\sum_l \cos{q_l}=0$\cite{muller-hartmann}.
Then,
the non--local parts of the second term in the RHS of Eq.~\ref{ward_dmfa} 
can be neglected, and the Ward identity, which now 
involves only the local $\Sigma ,\bgam $ and $G$ is exactly satisfied, 
as can be directly shown from the effective single site problem using 
equations of motion. 
However, there is a set of $\q$ of measure zero within the Brillouin
zone, which unfortunately includes the values $\q=0$
and $\q=(\pi,\pi,\ldots)$, for
which ${\bf X}(\q)$ is finite and 
$\chi_q^0(i\omega_n,i\nu_\alpha)\neq \chi_{ii}^0(i\omega_n,i\nu_\alpha)$,
with corrections of order unity.  For these values of $\q$ the
non--local parts in the second term can no longer be discarded, and the Ward
identity is again violated.  Consistent with this observation, one may 
show to all orders in perturbation theory that non-local 
corrections to the $D=\infty$ two--particle self energy remain finite 
for a set of measure zero points in the Brillouin zone.  Apparently, 
for these points, the non-local corrections 
to the two--particle self energy are needed to satisfy the Ward identity,
or, equivalently, the theory is only conserving if the limit as $D\to\infty$
is evaluated only after the functional derivatives of $\Phi$ (e.g.\
$\bgam_{\sigma,\sigma'}=\delta^2\Phi/\delta \G_\sigma\delta \G_{\sigma'}$) 
are evaluated\cite{caveat}.  

In a similar way, one may explore  violation of the Ward identities by the 
DCA. The required Ward identity in this case can be written as
\begin{eqnarray}
&&\Sigma _{c}({\bf K+Q},i\nu_{\alpha }+i\omega _{n})
-\Sigma _{c}({\bf K},i\omega_n) =  \\
&&\frac{T}{N}\sum_{\K', \tk, n'}[\{G(\K'+\tk,i\omega_n')
-G(\K'+ \tk+\Q+ \tq,i\omega _n'+i\nu _{\alpha })\} \nonumber \\
&&+\{\Sigma_c({\bf K}^{\prime}+{\bf Q},i\nu _{\alpha }+i\omega _{n})
-\Sigma _{c}({\bf K}^{\prime},i\omega _{n})\}
G(\K'+\tk,i\omega _{n}^{\prime})\times  \nonumber \\
&&G(\K'+\tk+\Q+\tq,i\omega _n'+i\nu_\alpha)]\times 
\mT^{(2)c}_{\Q+\tq, i\nu_\alpha}(\K, i\omega_n; \K', i\omega_n'),\nonumber
\end{eqnarray}
where we have used the DCA in assuming that $\Sigma $ and $\bgam $ are
dependent only on the cluster momenta, and $\T^{(2)c}$ is defined in
Appendix A.  Now it is clear that, to the extent
that the RHS depends on ${\bf {q}}$, the Ward identity will not be
satisfied, this even in the static case. 

However, the DCA will be conserving in the limit of large cluster size,
since momentum conservation at the internal vertices is restored
(with corrections of order $\Delta k$).
Here, we assume that the method used to solve the cluster is exact, 
or that if an approximate methods used, that the corresponding 
self-energy diagrams are formed from derivatives of a generating 
functional and employ fully dressed propagators 
(i.e., ${\bar{G}}(\k,\omega)$, not ${\cal{G}}(\k,\omega)$) so that
we approximate $\Phi(G(\k,\omega))\approx \Phi({{\bar{G}}}(\k,\omega))$.
Then, the DCA is conserving to the extent that $\bgam_\q(\k,\k')$ and 
$\Sigma_\k$ are well approximated by the cluster 
quantities.  Since ${\bgam}={\bgam}^c +{\cal{O}}(\Delta k^2)$ and 
$\Sigma=\Sigma^c +{\cal{O}}(\Delta k^2)$, the DCA is able to restore the
conservation properties lost in the DMFA when $\Delta k=\pi/L\to 0$
with corrections of order ${\cal{O}}(\Delta k^2)$. 

	In this Appendix we have shown that due to violations of
momentum conservation, the DMFA is not a conserving
approximation in any dimension $D$.  Violations of Ward's original identity 
also emerge for the DMFA even when $D\to\infty$ for a vanishingly small set 
of momenta $\q$ which includes  $\q=0$, but not for general 
momenta $\q$.   There are concomitant requisite non-local corrections to the 
infinite-dimensional irreducible vertex functions for a set of measure zero 
points in the infinite-dimensional Brillouin zone which are necessary to 
restore the Ward identity for all $\q$. In finite dimensions, the DMFA 
violates conservation in a finite fraction of the Brillouin zone due to the 
lack of momentum conservation in the internal vertices of the generating 
functional.  Momentum conservation is restored by DCA systematically as the
cluster size increases, and so the DCA restores the conserving nature of the 
approximation.


\begin{references}
\bibitem{choiceK} This is the only choice for the cluster $\K$ values
that are evenly spaced in the D-dimensional Brillioun zone and form a 
group under addition.  However, one may also choose $\K$ values that 
correspond to a cluster with lower dimensionality than $D$.  For example, 
one could choose just two cluster $\K$ values, corresponding to a two-site 
cluster.  However, care must be taken to properly symmetrize the corresponding 
cluster self energy, before using it in Eqs.~\ref{gbar} and \ref{gsck}.
\bibitem{average} In situations where only small clusters
are numerically accessible it might be advisable to average over
``all'' possible choices of the cluster momenta, thus reducing
the effect of ``special'' points in the Brillouin zone.
\bibitem{caveat} The fact that the DMFA is not conserving when $D\to\infty$
does not indicate that it is not exact in this limit.  In fact the theory
remains exact here, since the requisite non-local corrections to the 
irreducible vertex $\bgam$ occur only for a set of measure zero points 
in the Brillouin zone.  Thus, where $\bgam$ is used in an integral equation
to calculate a measureable quantity, such as a susceptibiltiy, or even
the DC conductivity, the contribution of these non-local corrections
is negligible.
\end{references}
\end{document}